\documentclass[times, review, 10pt]{elsarticle}

\usepackage{amsmath,amsfonts}
\usepackage{algorithm}
\usepackage{algpseudocode}
\usepackage{array}
\usepackage{graphicx}
\usepackage{float}
\usepackage{textcomp}
\usepackage{stfloats}
\usepackage{url}
\usepackage{verbatim}
\usepackage{amssymb}
\usepackage{tabularx}
\usepackage{multirow}
\usepackage{booktabs}
\usepackage[caption=false,font=footnotesize]{subfig}
\usepackage{bm}
\usepackage{makecell}
\usepackage{adjustbox}
\usepackage{appendix}
\usepackage{enumitem}
\usepackage{pifont}
\usepackage{pdfpages}
\usepackage{tikz}
\usepackage{tcolorbox}
\usepackage[normalem]{ulem}
\usepackage{xcolor}
\usepackage{hyperref}

\def\BibTeX{{\rm B\kern-.05em{\sc i\kern-.025em b}\kern-.08em
    T\kern-.1667em\lower.7ex\hbox{E}\kern-.125emX}}

\definecolor{red}{HTML}{D00004}
\newcommand{\red}[1]{\textcolor{red}{#1}}
\long\def\comment#1{}

\def\ie{$i.e.$}
\def\eg{$e.g.$}

\newcommand{\blue}[1]{\textcolor{blue}{#1}}

\journal{Pattern Recognition}

\begin{document}

\begin{frontmatter}

\title{Coward: Collision-Based OOD Watermarking for Practical Proactive Federated Backdoor Detection}

\author[tsinghua]{Wenjie Li}

\author[ecnu]{Siying Gu}

\author[ntu]{Yiming Li}

\author[ntu]{Shuxin Li}

\author[ecnu]{Zhili Chen}

\author[ntu]{Tianwei Zhang}

\author[tsinghua]{Shu-Tao Xia}



\affiliation[tsinghua]{
  organization={Tsinghua University},
  country={China}
}

\affiliation[ecnu]{
  organization={East China Normal University},
  country={China}
}

\affiliation[ntu]{
  organization={Nanyang Technological University},
  country={Singapore}
}

\begin{abstract}
Backdoor detection is currently the mainstream defense against backdoor attacks in federated learning (FL), where a small number of malicious clients can upload poisoned updates to compromise the federated global model. Existing backdoor detection techniques fall into two categories, passive and proactive, depending on whether the server proactively intervenes in the training process. However, both of them have practical limitations: passive detection methods are disrupted by common non-i.i.d. data distributions and random participation of FL clients, whereas current proactive detection methods are misled by an inevitable out-of-distribution (OOD) bias because they rely on \emph{backdoor coexistence effects}. To address these issues, we introduce a novel proactive detection method dubbed \texttt{Coward}, inspired by our discovery of \emph{multi-backdoor collision effects}, in which consecutively planted, distinct backdoors significantly suppress earlier ones. Correspondingly, we modify the federated global model by injecting a carefully designed backdoor-collided watermark, implemented via regulated dual-mapping learning on OOD data. This design not only enables an inverted detection paradigm compared to existing proactive methods, thereby naturally counteracting the adverse impact of OOD prediction bias, but also introduces a low-disruptive training intervention that inherently limits the strength of OOD bias, leading to significantly fewer misjudgments. Extensive experiments on benchmark datasets show that \texttt{Coward} achieves state-of-the-art performance and effectively alleviates OOD bias.
\end{abstract}



\end{frontmatter}

\section{Introduction}
Federated Learning (FL)~\cite{mcmahan2017fedavg} has emerged as a powerful paradigm for privacy-preserving machine learning, enabling distributed clients to collaboratively train models without sharing raw data, and is becoming a promising basis for trustworthy pattern recognition applications. By keeping data locally owned and processed, FL inherently strengthens privacy and copyright protection, and unlocks the value of proprietary domain-specific and personal data held by private organizations and individuals. This privacy-native property makes FL particularly suitable for building secure and personalized intelligence that goes beyond models trained solely on generic public data~\cite{pr2026_ppfl_he,pr2024_fzsl,pr2026_fedofa}.
As a result, FL has been extensively studied across many pattern recognition domains, such as healthcare and finance~\cite{pr2023_fedcl, pr2024_med_survey, fl_finance_icse24}.

\begin{figure}[t]
    \centering
    \includegraphics[width=0.8\linewidth]{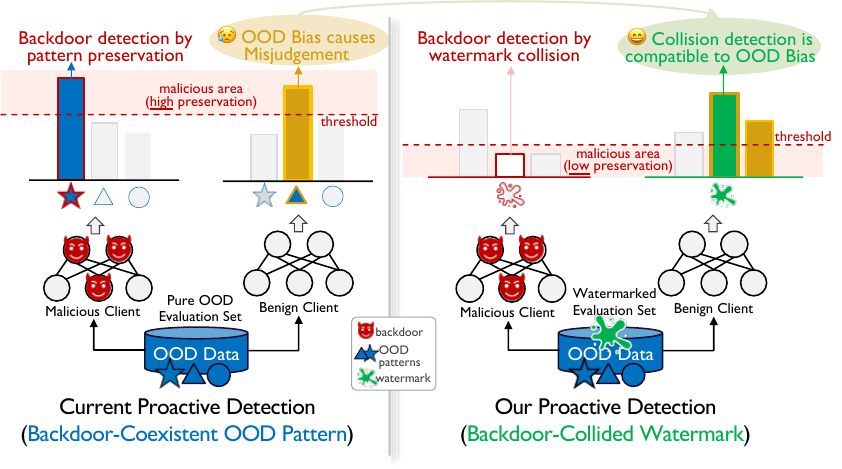}
    \vspace{-1em}
    \caption{\textbf{Our backdoor-collided detection versus the existing backdoor-coexistent method}. With our backdoor-collided watermark, we enable an inverted detection paradigm that effectively mitigates misleading effects caused by the OOD prediction bias, a key challenge that limits current proactive detection methods.}

    \label{fig:fig_intro}
\end{figure}

However, the privacy-preserving design of federated learning is a double-edged sword. While keeping raw data local effectively protects sensitive information, it also restricts the server’s ability to inspect participant behavior during training. This restricted observability creates an inevitable vulnerability to model-level attacks, where the integrity of the federated global model can be compromised by covertly manipulated local models, most prominently through backdoor attacks~\cite{ba_survey_yiming, chameleon,pr2026_rethinking_trigger}. In general, backdoor attacks cause models to behave normally on benign inputs while producing harmful outputs only under specific trigger conditions, rendering the malicious behavior stealthy and difficult to detect. Once the malicious models are aggregated together with benign ones, the backdoor is seamlessly propagated to the global model, thereby manipulating the entire federated system~\cite{bagdasaryan2020backdoor,pr2026_mtfba}.

To detect and exclude malicious backdoor models, passive detection methods have been proposed~\cite{rieger2022deepsight, nguyen2022flame}, which do not intervene in the FL process and assume that malicious models manifest observable model-level deviations as outliers. They rely on similarity-based measurements over either model parameters or outputs and identify anomalous models either through direct thresholding or by further applying clustering to expose structurally isolated outliers. However, we reveal that these methods are vulnerable to client non-i.i.d.\ data distributions and participation randomness, which are typical in FL~\cite{mcmahan2017fedavg}. These factors lead to highly dynamic and divergent benign local models that may appear as outliers, significantly interfering with detection. We refer to this limitation as the \textbf{non-i.i.d. divergence challenge}, as analyzed in Section \ref{sec:passive_revisit}.

Recently, the pioneering work BackdoorIndicator~\cite{li2024backdoorindicator} proposed the first federated \emph{proactive detection} strategy, where the server explicitly intervenes during training to induce malicious models to exhibit behaviors distinct from benign ones. Specifically, BackdoorIndicator injects OOD patterns designed to be easily retained by malicious models while tending to be forgotten by benign ones, such that stronger pattern retention in a local model indicates a malicious client. By \textit{shifting the detection basis from model discrepancies to differences in client responses to intervention actions}, proactive methods naturally avoid the sensitivity to non-i.i.d. distributions that often undermines passive detection methods. However, due to the privacy constraints of FL that prevent access to in-distribution client data, proactive training interventions at the server are restricted to operating via OOD data. As a result, detection outcomes become susceptible to the OOD prediction bias of deep neural networks (DNNs)~\cite{nguyen2015ood_bias,hein2019relu_ood_bias_theory}. This bias causes benign local models to be easily misidentified as malicious clients. We provide a detailed analysis in Section~\ref{sec:proactive_revisit} and refer to this limitation as the \textbf{OOD bias challenge}. 

To tackle the above challenges, we propose \underline{C}ollision-based \underline{O}OD \underline{WA}termarking for \underline{R}obust federated backdoor \underline{D}etection, termed \texttt{Coward}. Our method is inspired by the finding of the \emph{multi-backdoor collision effect}, where a consecutively planted distinct backdoor significantly interferes with the one previously planted. To this end, we detect backdoor models by evaluating whether the server-injected conflicting global watermark has been erased during local training. 
Concretely, the global OOD watermark consists of a regulated base OOD mapping and a targeted watermark mapping, where the former assigns a static correspondence from the OOD sample space to the task label space, and the latter connects triggered OOD samples to the watermark target class. In general, our method alleviates the OOD bias challenge through three aspects: \textbf{1)} Unlike prior approaches, our collision-based detection treats high inspection accuracy as the signal of benign behavior, making it naturally compatible with the OOD bias–induced high-confidence predictions; \textbf{2)} Our watermark-based detection inspects the presence of the trigger-to-target mapping rather than the OOD sample-to-target mapping, mitigating the direct impact of OOD prediction bias. \textbf{3)} Our method enforces explicit class-wise label assignments to OOD samples to reduce random prediction fluctuations. We demonstrate (1) and (2) in Figure~\ref{fig:fig_intro}, and verify (3) in Figure~\ref{fig:overall_ood_bias_evidence}.

Our main contributions are summarized as follows: \textbf{(1)} We revisit existing detection methods for FL backdoor attacks and identify two key challenges: non-i.i.d. divergence and OOD prediction bias. We show how these challenges undermine existing detection methods, leading to high benign-client misclassification. \textbf{(2)} We find an intriguing phenomenon, dubbed the multi-backdoor collision effect, where a consecutively planted distinct backdoor can significantly suppress the previously implanted one. Based on this finding, we design a simple yet effective proactive backdoor detection method, \ie, \texttt{Coward}, that effectively mitigates OOD bias and achieves more accurate detection. \textbf{(3)} Extensive experiments on benchmark datasets verify the effectiveness of our method and its resistance to potential adaptive backdoor attacks.

\section{Background and Related Works}
\label{sec:relatedworks}

\subsection{Federated Learning}
%

Federated learning enables multiple decentralized clients to collaboratively train a shared model without exchanging their private data. Based on how data is partitioned, FL is typically categorized into horizontal FL (HFL), where clients share the same feature space but different samples, and vertical FL (VFL), where clients share samples but differ in input fields~\cite{flsurvey_yangqiang}. In this paper, we focus on the HFL setting, where a central server coordinates model training by periodically selecting a subset of clients to receive the current global model, update it using local data, and return the updated parameters. The server then aggregates these updates to form a new global model, and this process repeats iteratively, progressively improving model performance while keeping all training data local. Specifically, given the selected client set $\mathcal{S}^{t-1}$ in round $t-1$, the global model $\bm{w}^t$ in round $t$ is:
\begin{equation}
\bm{w}^t = \bm{w}^{t-1} + \sum_{k\in \mathcal{S}^{t-1}} \frac{N_k}{N}\Delta\bm{w}^{t-1}_k,
\end{equation}
where $N_k$ and $N$ denote the data volume of client $k$ and the total data volume across participating clients, respectively. $\Delta$ denotes the model residual with respect to the previous global model. We highlight three key characteristics that distinguish FL from centralized settings: \textbf{(1) Multi-party Participation}: The global model is jointly trained by multiple clients, each of which contributes to the model privately but cannot individually control the training outcome. \textbf{(2) Non-i.i.d. Client Data} \cite{mcmahan2017fedavg}: Data distributions can vary significantly across clients, leading to divergent local updates and making the global optimization layer more challenging. \textbf{(3) Partial Client Participation}\cite{mcmahan2017fedavg}: In each round, only a subset of clients is randomly selected due to limited bandwidth and dynamic client availability, increasing the uncertainty of training dynamics.


\subsection{Backdoor Attacks in Federated Learning}
Backdoor attacks aim to manipulate neural networks to function normally on clean data but misclassify inputs containing attacker-specified triggers~\cite{ba_survey_yiming,gu2020badnets}.
Data poisoning is the most common backdoor injection method, where a subset of training samples is embedded with triggers and mislabeled to a target class~\cite{gu2020badnets}. Stealth and persistence are key objectives in backdoor attacks, motivating the design of invisible~\cite{nguyen2021wanet,Gao_attack24}, semantic~\cite{xu2024csba_semantic_atk}, optimized~\cite{sun2024backdoor}, sample-specific~\cite{zhu2025towards} and transformation-aware~\cite{pr2026_rethinking_trigger} backdoor strategies.
However, these approaches are limited in FL, because adversaries can only control a limited \emph{subset of clients} and must ensure their backdoored updates \emph{survive the aggregation process} to impact the global model.
Specifically, FL-specific backdoor attacks can be classified into the following categories.

\vspace{0.2em}
\noindent \textbf{Single-client Attack.} The first backdoor attack method for FL introduced the model replacement attack technique~\cite{bagdasaryan2020backdoor}, where a malicious client uploaded a scaled local update to dominate the aggregated model. However, the abnormally large update made it easily detectable by detection methods~\cite{zitengsun01pgd}. To improve \textit{stealth} and \textit{persistence}, follow-up works introduced more subtle attack strategies. For example, PGD~\cite{zitengsun01pgd} constrained malicious updates within an $\ell_2$-ball to evade norm-based detection, while Neurotoxin~\cite{Zhengming01Neurotoxin} injected backdoors into parameters associated with inactive neurons to minimize noticeable changes. Other works aimed to further blend backdoor patterns into normal behavior. For example, Edge-case attack~\cite{wang2020edgecase} utilized naturally occurring features as triggers, reducing the likelihood of detection, while Chameleon~\cite{chameleon} aligned backdoor and benign features using contrastive learning.


\vspace{0.2em}
\noindent \textbf{Multi-client Attack.} The DBA framework~\cite{xie2019dba} leveraged the distributed nature of FL to spread the backdoor across multiple clients, with each client using a distinct local trigger. These independently subtle triggers collectively formed an effective global backdoor, reducing the detectability of any individual client. Expanding on this idea, collusive multi-client attacks~\cite{gong2022coordinated_dba,pr2026_mtfba} have been proposed, in which backdoor triggers are further optimized in a model-aware manner. Such learning-based coordination enables attackers to craft triggers that are both less suspicious and more effective, thereby amplifying the threat posed by distributed poisoning. In contrast, Non-Cooperative Backdoor Attacks ~\cite{nguyen2024nba} involved independent attackers using distinct triggers for different targets, yet still exhibited enhanced stealth comparable to coordinated methods. 

\subsection{Backdoor Defenses in Federated Learning}
In centralized settings, backdoor defenses typically focus on either removing the backdoor from the model~\cite{wu2021ba_defense_prune,pr2026_appckd_defense,pr2026_bam_defense} or filtering trigger samples during training or inference~\cite{qiu2021deepsweep,pr2026_pdi_defense}. However, these approaches are often \emph{impractical in FL due to the defender's limited capabilities} (\ie, the central server): \textbf{(1)} the server lacks access to clean and in-distribution data, making data-dependent removal methods like pruning, retraining, or fine-tuning infeasible. \textbf{(2)} the server cannot control the local training process, rendering sample filtering and robust training strategies infeasible. To address these limitations, FL-specific backdoor detection mechanisms have been developed, generally falling into the following two categories.

\vspace{0.2em}
\noindent \textbf{Backdoor Effect Mitigation} methods aim to reduce the impact of backdoor behaviors in the global model by suppressing anomalous neurons across local models. Early works such as Weak-DP~\cite{zitengsun01pgd} applied norm clipping and noise injection to weaken abnormal neuron effects. Statistical approaches like geometric median~\cite{RFA} and Trimmed-Mean~\cite{blanchard2017machine} enforced robustness by aggregating updates in a statistically neutral manner. RLR~\cite{ozdayi2021RLR} further enhanced mitigation granularity by assigning parameter-wise adaptive learning rates. FLPurifier~\cite{zhang2024flpurifier} achieved stronger backdoor suppression by performing adaptive aggregation at the classifier level under a modified FL protocol that decoupled encoder and classifier training. However, these methods \textit{operate all clients equally, often diminishing benign model contributions and harming overall performance} under the standard FL protocol. This limitation has led to the rise of the following detection-based methods as a mainstream alternative.

\vspace{0.2em}
\noindent{\textbf{Backdoor Client Detection}} methods aim to identify and exclude malicious local models during aggregation, typically by quantifying distance between benign and malicious client models to find outliers. For instance, Multi-Krum~\cite{multiKrum} leveraged Euclidean distance, while Foolsgold~\cite{foolsgold}, FLTrust~\cite{cao2021fltrust}, and ShieldFL~\cite{ma2022shieldfl} adopted cosine similarity, all applied directly on full model weights. Rflbat~\cite{wang2022rflbat} extended this by computing distances in a top-$k$ PCA-projected space, aiming to reduce noise and improve efficiency, though at the cost of some information loss. Beyond model parameters, DeepSight~\cite{rieger2022deepsight} incorporated prediction-level discrepancies of local models, enabling a more comprehensive analysis. Extending beyond pure detection, Flame~\cite{nguyen2022flame} combined clustering-based client identification with gradient norm-based mitigation, positioning itself at the intersection of detection and mitigation strategies and demonstrating strong empirical performance. FLGuardian~\cite{tifs25_fl_defens_poison}, a recent study addressing untagged backdoor attack, performed layer-wise model analysis using similarity metrics and clustering algorithms, reflecting a growing shift toward more fine-grained detection in the parameter space. However, these methods \textit{assume malicious models are outliers, making them sensitive to data heterogeneity and ineffective against stealthy attacks}.

Since all the above detection methods operate without any server-side intervention, we categorize them as \textit{passive detection}. In contrast, \textit{proactive detection} explicitly intervenes in FL training to facilitate backdoor identification. The recent pioneering work Indicator~\cite{li2024backdoorindicator} adopts this strategy by injecting OOD mappings into the global model, achieving improved robustness under data heterogeneity. However, it still suffers from high false-positive rates due to OOD prediction bias. Recent work~\cite{tdsc25_fba_dfs2_proactive_poi} has also applied the proactive detection paradigm to poisoning attacks by amplifying directional deviations between malicious and benign updates through broadcasting a scaled and perturbed, detection-oriented surrogate of aggregated gradients. However, this approach relies on the explicit accuracy-degradation objective of poisoning attacks, which is less stealthy than backdoors and easier to detect, and further depends on parameter-space gradient comparisons that are susceptible to non-i.i.d. bias. In summary, effective proactive backdoor detection in FL remains an open challenge.

\section{Revisiting Existing Detection Methods and Multi-Backdoor Interplay}
\label{sec:revisiting}

In this section, we first introduce the threat model and revisit the limitations of existing detection methods. We then examine the interplay effects of multi-backdoor attacks, which directly motivate our detection method.

\subsection{Preliminary: Threat Model}

\vspace{0.3em}
\noindent \textbf{Attacker's Goal and Capability.} The attacker resides on the client side and aims to inject a targeted backdoor into the global model, causing it to misclassify all trigger-embedded inputs into a designated target class while maintaining normal performance on benign inputs.
Although the attacker cannot directly access the global model, they may control a subset of clients and submit arbitrary malicious models or gradients to the server, thereby influencing the global model updates.

\vspace{0.3em}
\noindent \textbf{Defender's Goal and Capability.} The defender is the central server, whose goal is to identify and exclude backdoored local models from the aggregation process. The server cannot access any data from the clients' local distributions but has full visibility into all uploaded local models. It also controls the update and broadcast of the global model. Additionally, the server is allowed to make arbitrary modifications to the global model and has access to any out-of-distribution data. 

\subsection{Revisiting Passive Detection}
\label{sec:passive_revisit}
Many existing passive detection methods assume that \textit{malicious updates deviate from benign updates in the parameter space} due to the additional optimization of the backdoor objective during local training~\cite{blanchard2017machine}. Based on this assumption, many methods leveraging clustering and outlier filtering have been proposed to detect malicious clients~\cite{RFA, nguyen2022flame}. To validate this assumption under practical non-i.i.d. settings, we analyze the statistical characteristics of local model updates and show their correlation with misjudgments of benign clients.

\vspace{0.3em}
\noindent \textbf{Settings.} We revisit the gradient norm distribution for each client under a highly non-i.i.d. partition of the CIFAR10 dataset, with label distribution generated by a Dirichlet prior with $\alpha=0.3$, and participation by 10 randomly clients out of 100. One malicious client performs a backdoor attack using a pixel-style trigger targeting class 0.

\begin{figure}[t]
    \centering
    \subfloat[Non-i.i.d. Distraction Analysis]{%
        \includegraphics[width=0.25\linewidth]{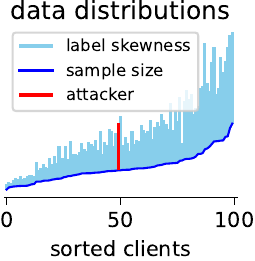}
        \includegraphics[width=0.25\linewidth]{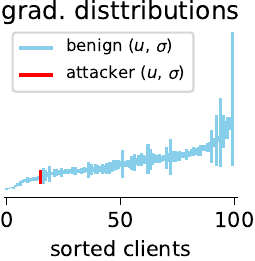}
        \includegraphics[width=0.25\linewidth]{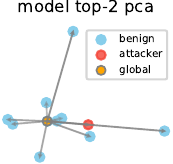}
    }\\

    \subfloat[Norm-Misjudgment Correlation Analysis]{
        \includegraphics[width=0.75\linewidth]{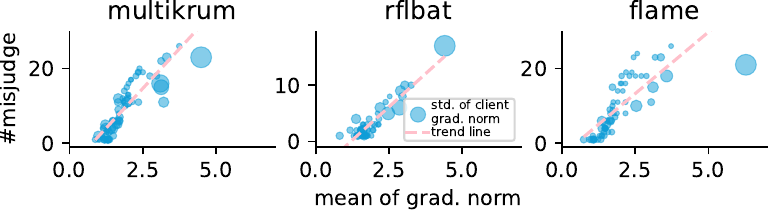}
    }
    \caption{\textbf{The distraction effect of non-i.i.d. data on passive detection methods.} Divergent client data distributions (left of Figure (a)) substantially reduce the suspiciousness of malicious clients, as reflected in both gradient norms (middle) and the update directions of benign models (right). In contrast, it increases the perceived suspiciousness of benign clients, particularly those with larger gradient norms, as evidenced by the positive correlation observed in Figure (b).}
    
    \label{fig:non_iid_bias}
\end{figure}

\vspace{0.3em}
\noindent \textbf{Results.} As shown in Figure~\ref{fig:non_iid_bias}(a), \textit{the suspicion toward attack is significantly reduced due to the severity of data heterogeneity}: {\textbf{(1)}} in the first column, we present the size and label skewness of local datasets. The blue line indicates the dataset size, while the bar height represents the skewness, measured by the standard deviation of class-wise sample counts. We can observe that client data distributions are highly diverse. Notably, malicious clients do not exhibit abnormal characteristics in either size or skewness; their distributions fall within a typical range. In contrast, some benign clients show extreme distributions with both large data volumes and high skewness. {\textbf{(2)}} In the second column, we plot the mean and standard deviation of gradient norms uploaded by all 100 clients across training rounds, sorted by their mean values. The results reveal that the malicious client's updates have slightly smaller gradient norms compared to most benign clients and appear inconspicuous amid the naturally diverse benign client updates. {\textbf{(3)}} In the third column, we visualize the top-2 PCA component of local and global models (based on the last-layer weights) during training. The results show that the backdoor models' update direction and distance lie in a moderate region, appearing unharmful due to the diverse update directions introduced by the non-i.i.d. distribution. Figure~\ref{fig:non_iid_bias}(b) further presents a correlation analysis, \textit{revealing a clear association between model update norms and misjudgment of malicious clients}. Specifically, the x-axis shows the mean gradient norm, the y-axis indicates the number of misjudgments, and the bubble size reflects the standard deviation of the gradient norm.
The upward trend suggests that clients with larger and more variable gradient norms are more susceptible to being misjudged. In conclusion: 
\begin{tcolorbox}[colback=blue!2!white,leftrule=1mm,size=title]
    \emph{\textbf{Takeaway 1}: Data heterogeneity (\ie, non-i.i.d.) amplifies client model variance, blurring distinctions between malicious and benign clients and diminishing the effectiveness of passive backdoor detection.}
\end{tcolorbox}
To mitigate this limitation, proactive methods decouple detection from parameter differences and instead rely on client-side reactions to server-side intervention actions, which effectively alleviates sensitivity to data heterogeneity.

\begin{figure}[!t]
    \centering
    \includegraphics[width=0.85\linewidth]{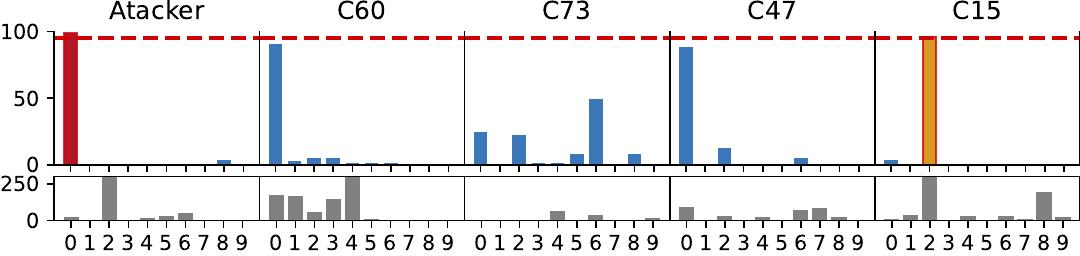}
    
    \begin{tikzpicture}
    \draw[dashed, thick] (0,0) -- (0.85\linewidth,0) node[midway, above] {$\uparrow$ \textit{w./o. planting} $\uparrow$} node[midway, below] {$\downarrow$ \textit{w./ planting} $\downarrow$};
    \end{tikzpicture}
    
    \includegraphics[width=0.85\linewidth]{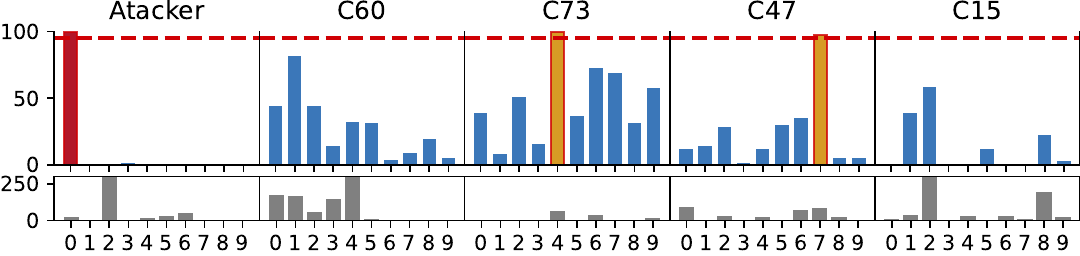}
    \vspace{-1em}
    \caption{\textbf{The misdirection effect of OOD bias against the existing proactive detection methods.} With the attacker-specified target class set to `0' (\red{red}), we show 5 clients from the same round, presenting their OOD prediction distribution, with results in the no-planting case in the first row and the planting case in the second row. The gray subfigure shows the local data distribution. The red dashed line marks the detection threshold; classes with inspection accuracy above this line are flagged as malicious and highlighted in red border. The results indicate that: (1) Even without server-side planting, non-target classes may exhibit OOD-induced high inspection accuracy (\textcolor{orange}{orange}) on benign clients; (2) With planting, misjudgment increases, and more classes exhibit higher inspection accuracy.}
    \label{fig:ood_bias_detail}

\end{figure}

\begin{figure}[t]
    \centering
    \subfloat[Client-wise OOD Bias]{%
        \includegraphics[width=0.24\linewidth]{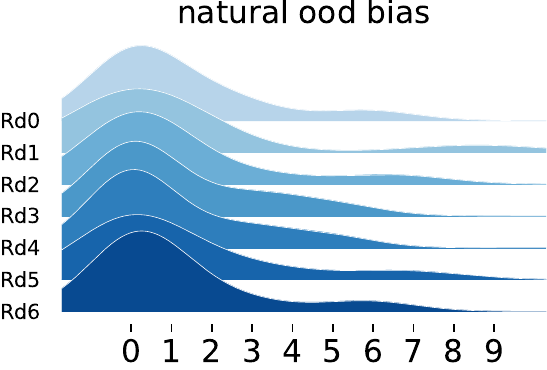}
        \includegraphics[width=0.24\linewidth]{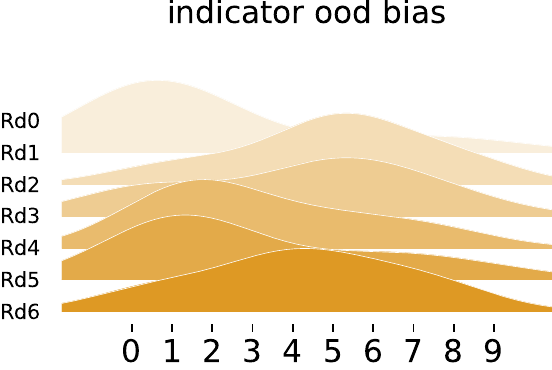}
        \label{fig:natural_bias}
    }\hfill
    \subfloat[Overall OOD Bias.]{%
        \includegraphics[width=0.48\linewidth]{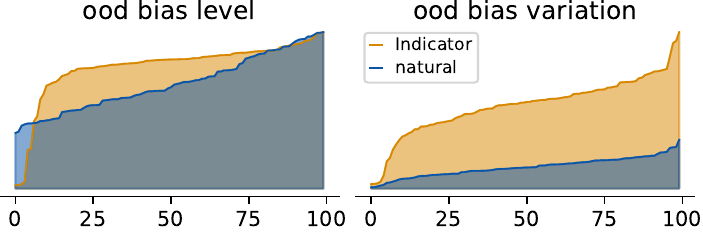}
        \label{fig:overall_bias}
    }
    \caption{\textbf{Exacerbation of OOD bias induced by existing proactive methods}. The Indicator method intensifies both the magnitude and variation of OOD bias.}
    \label{fig:ood_bias_excabration}
\end{figure}

\subsection{Revisiting Proactive Detection}
\label{sec:proactive_revisit}

As a pioneering work in proactive detection, Indicator identified a \textit{co-existing} (\ie, maintaining) \textit{effect}, where a subsequently planted backdoor with the same target label could preserve the effect of a previously injected one.
However, due to the privacy constraint of federated learning, it could not access client data and had to rely solely on OOD data to modify the global model.
To address this constraint, Indicator implanted \textit{random OOD mappings} (\ie, assigning labels to OOD samples via uniform random sampling) into the global model to activate the co-existing effect.
During detection, it evaluated each client's class-wise prediction accuracy on these OOD data: any class exceeding a predefined threshold flagged the client as malicious, and the corresponding class was assumed to be the attacker's backdoor target.
Such a detection mechanism no longer relies on parameter space, making it less susceptible to the distraction effect of non-i.i.d. data and thus more effective than passive detection methods.

However, leveraging OOD samples to activate the co-existing effect introduces a vulnerability to \textit{OOD prediction bias}, where deep neural networks inevitably tend to classify OOD samples into arbitrary classes with high confidence~\cite{nguyen2015ood_bias, hein2019relu_ood_bias_theory}. Specifically, it leads to a high false positive rate, as analyzed in the following parts.

\vspace{0.3em}
\noindent \textbf{Settings.} We analyze the prediction distribution of OOD samples under Indicator~\cite{li2024backdoorindicator}, both with and without server-side intervention. The main FL task is CIFAR-10 classification, with EMNIST used for proactive pattern planting. During planting, Indicator randomly partitioned OOD samples in the label space with the same 10\% ratio. In each round, 10 clients are sampled from a total of 100, with one attacker targeting class 0. The data heterogeneity is controlled by a Dirichlet distribution with $\alpha = 0.3$.

\vspace{0.3em}
\noindent \textbf{Results.} As shown in Figure~\ref{fig:ood_bias_detail}, we find that: \textbf{(1)} under the no-intervention setting, for the malicious client, OOD samples successfully activate the learned backdoor pattern with high accuracy, revealing an empirical correlation between the backdoor trigger and the OOD pattern. Meanwhile, for benign clients, certain classes exhibit inspection accuracy that approaches or even exceeds the predefined detection threshold of 95\%, resulting in significant misjudgment; \textbf{(2)} Under the with-intervention setting, the pre-planted OOD pattern leads to similar effects and, in some cases, causes even more classes to maintain higher accuracy among benign clients (\ie, tending to be treated as malicious clients). These phenomena indicate that:

\begin{tcolorbox}[colback=blue!2!white,leftrule=1mm,size=title]
    \emph{\textbf{Takeaway 2}: OOD prediction bias is an inherent vulnerability of DNNs~\cite{nguyen2015ood_bias, hein2019relu_ood_bias_theory}, causing OOD samples to be classified into arbitrary classes with high confidence. Under Indicator's detection scheme, this bias can give benign clients unexpectedly high suspicion scores, leading to frequent yet unpredictable false positives.}
\end{tcolorbox}

Going further, we hereby demonstrate how the intervention action intensifies OOD prediction bias in Figure~\ref{fig:ood_bias_excabration}, from which we draw the following key observations. \textbf{{(1)}} \textit{Disruption of temporal consistency}: In subfigure (a), we present the prediction distributions of a specific client across multiple global rounds.
The no-intervention setting maintains a stable bias toward a particular class over time, whereas the with-intervention setting exhibits a highly variable trend, \ie, the class with the highest prediction accuracy fluctuates significantly across rounds.
This instability is likely caused by the random label assignment used in the Indicator method, where OOD samples with similar semantic content may be mapped to different labels during the intervened training process, creating a disordered association between the OOD space and the task label space; \textbf{{(2)}} \textit{Amplified bias level and variation}: As shown in subfigure (b), we quantify the overall bias level and its temporal variation across all 100 clients, sorted by bias magnitude.
The results show that Indicator's intervention introduces significantly higher bias intensity and instability compared to the no-intervention case.
We further analyze the correlation between OOD bias levels and false positive rates in Section \ref{sec:experimenal}(Figure~\ref{fig:overall_ood_bias_evidence}), showing a strong correlation.
To this end, we conclude that:
\begin{tcolorbox}[colback=blue!2!white,leftrule=1mm,size=title]
    \emph{\textbf{Takeaway 3}: BackdoorIndicator exacerbates OOD prediction bias through random mapping planting, which not only disrupts bias patterns but also introduces temporal inconsistency across rounds, ultimately undermining its reliability.}
\end{tcolorbox}

Given the limitations of random OOD planting, the effectiveness of leveraging the multi-backdoor coexistence effect behind Indicator becomes questionable. This naturally leads to a deeper question: \textit{Does an alternative multi-backdoor interaction mechanism exist, and if so, can it offer a more robust foundation for utilizing OOD data?} Accordingly, we shift to revisit multi-backdoor interplay.

\subsection{Revisiting Multi-Backdoor Interplay}
\label{subsec:revisit_collision}

Unlike Indicator's observation of a co-existence effect between backdoors targeting the same label, we identify a distinct \textit{collision effect}, where sequentially implanted backdoors targeting different labels tend to interfere with one another. 

\begin{figure}[t]
    \centering
    \includegraphics[width=0.35\linewidth]{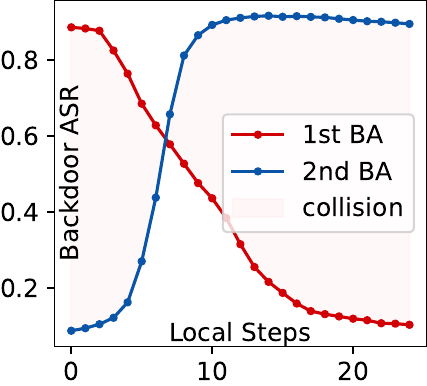}
    \includegraphics[width=0.35\linewidth]{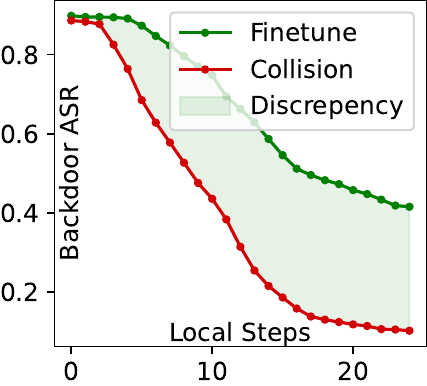}
    \vspace{-1em}
    \caption{\textbf{The collision effect between backdoors with different targets}. \textbf{(Left)}: A subsequently injected backdoor significantly degrades the ASR of the first backdoor. \textbf{(Right)}: The ASR degradation of the first backdoor caused by standard fine-tuning is notably smaller than that caused by subsequent backdoor injection. Together, they clearly demonstrate the presence of a collision effect.}
    \label{fig:collision}
\end{figure}

\vspace{0.3em}
\noindent \textbf{The Demonstration of Collision Effect}. We conduct the experiment under a centralized setting simulating two entities sequentially planting backdoors into the same ResNet model on the CIFAR-10 dataset but with different target labels. Specifically, the first backdoor uses a WaNet trigger targeting class 1 while the second backdoor uses a pixel-pattern trigger targeting class 0. Following \cite{li2024backdoorindicator}, the batch normalization statistics are also switched accordingly. As shown in Figure~\ref{fig:collision}, the left subfigure demonstrates that the ASR of the first backdoor steadily decreases as the second backdoor is progressively injected. In contrast, the right subfigure shows that the ASR decline from normal fine-tuning (due to knowledge forgetting) remains clearly less severe, highlighting the distinct collision effect between the two backdoors.

\vspace{0.3em}
\noindent \textbf{Collision Effect vs. Coexistence Effect.} 
To this end, we can obtain a complete understanding of multi-backdoor interplay under sequential implantation: Backdoors targeting the same class tend to coexist, while those with different targets often interfere with each other.
Though seemingly opposite, the two mechanisms can be understood through a unified and intuitive perspective.
As noted by Indicator~\cite{li2024backdoorindicator}, backdoor triggers can be interpreted as OOD features in the benign latent space, typically residing in a shared OOD region within the latent space.
When multiple triggers share the same target label, they construct consistent mappings from the latent OOD region to that label.
This alignment allows backdoors to reinforce each other, resulting in their coexistence.
In contrast, when triggers are associated with different target labels, the newly implanted trigger disrupts the existing mapping from the latent OOD region to the prior label, leading to collision or substitution.

\vspace{0.3em}
\noindent \textbf{Insights from Collision Effect.} Based on the collision effect, it becomes feasible to design a backdoor-based watermark that intentionally interferes with the attacker's trigger while remaining intact in benign clients. This leads to a new proactive detection paradigm distinct from the existing coexistence-based methods, where client models that fail to retain the watermark (\ie, exhibit low inspection accuracy) are identified as malicious, while those that preserve it (\ie, show high inspection accuracy) are considered benign. Under this inverted paradigm, the high inspection accuracy induced by OOD prediction bias in benign clients no longer misleads detection; instead, it becomes beneficial in some cases. Building on these insights, we propose our collision-based watermarking method, as detailed in the next section.

\section{Methodology}
\label{sec:methods}

\begin{figure*}[t]
    \centering
    \includegraphics[width=1\linewidth]{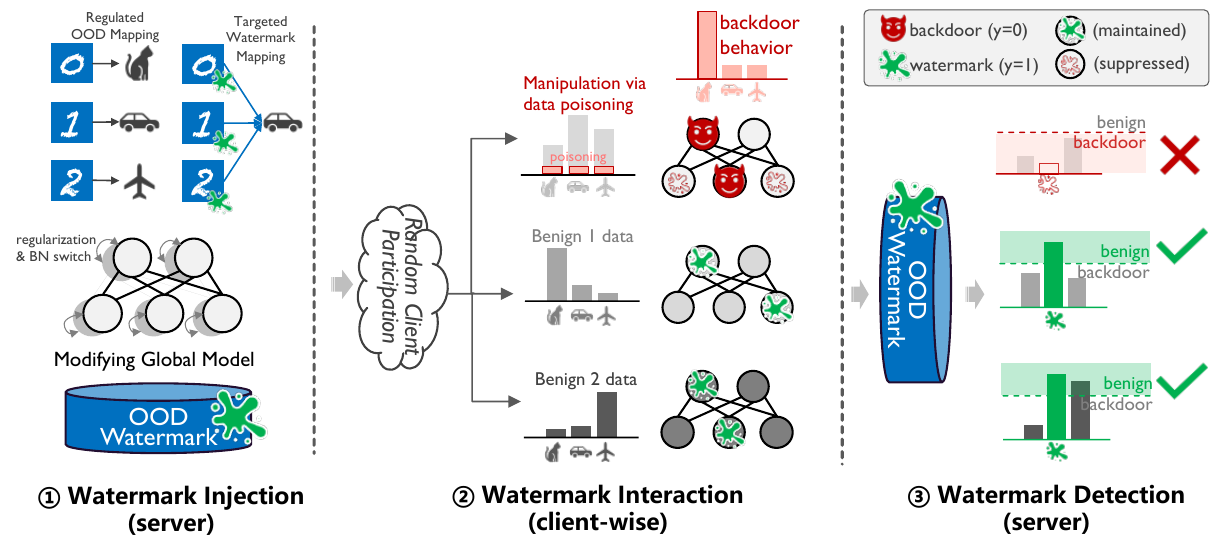}
    \vspace{-2em}
    \caption{\textbf{The overall pipeline of our \texttt{Coward} method}. \textbf{(left)} A forensic watermark is carefully embedded into the global model via low-cost OOD data training. \textbf{(Middle)} Random participants conduct local training based on the watermarked global model, where attackers tend to remove the watermark while benign clients preserve it. \textbf{(Right)} After local training, the server inspects the strength of the watermark; models with diminished watermark signals are flagged as malicious.}
    \label{fig:pipeline}
\end{figure*}

Figure~\ref{fig:pipeline} outlines three stages of our method: watermark injection, watermark interaction, and watermark detection. We elaborate on each stage below.

\subsection{Watermark Injection}
Watermarking is common for integrity protection by embedding identifiable patterns into models or data \cite{liu2024wm_llm_survey}. Inspired by this idea and our observation of the backdoor collision effect, we leverage a backdoor-based watermarking strategy to identify malicious clients in FL. To adapt the idea of backdoor watermarking to the out-of-distribution setting, we first revisit the conceptual structure of traditional backdoor mechanisms. In conventional in-distribution scenarios, a backdoor can be viewed as composed of two parts: \textbf{(1)} a standard task mapping that aligns clean inputs with their ground-truth labels, and \textbf{(2)} a tampered mapping that redirects triggered inputs to attacker-specified labels. Following this intuition, we borrow a similar two-part structure in the OOD context: an \textit{OOD base mapping} for clean OOD samples and an \textit{OOD watermark mapping} for triggered ones. 

\vspace{0.3em}
\noindent \textbf{Planting a Regulated Base Mapping.}
In FL, the OOD planting set and client training data share the same global model, thus operating within the same prediction space. However, their ground-truth semantic spaces are inherently different. This misalignment introduces \textit{prediction ambiguity}, where the same output value may imply conflicting meanings. For instance, label “0” might refer to the digit “0” in an EMNIST-based OOD planting set, but to “airplane” in a CIFAR-10 client task. Such ambiguity can lead to two negative effects: \textbf{(1)} the OOD mapping may distort client-learned semantics, reducing benign accuracy; \textbf{(2)} the non-i.i.d. nature of client training may corrupt the OOD mapping.
To address these problems, we manually assign a fixed one-to-one label mapping $\pi_1$ from the OOD label space $\mathcal{Y}_o$ to the task label space $\mathcal{Y}_t$. This defines the base mapping planting:
\begin{align}
    &\pi_1 : \mathcal{Y}_o\rightarrow\mathcal{Y}_t, \quad
    \mathcal{D}_p = \{ (\bm{x}_i, \pi_1(y_i)) \mid y_i \sim \mathcal{Y}_o \}, \\
    &\mathcal{L}_{base} = \sum_{i\in\mathcal{D}_p} \mathrm{CE}(f(\bm{x}_i; \bm{w}), \pi_1(y_i)).
\end{align}
Here, $\mathcal{D}_p$ denotes the planting set, $\bm{w}$ represents the model parameters, and $\mathrm{CE}$ is the cross-entropy loss. Since OOD samples lack explicit semantic grounding in the task label space, the mapping's meaning is not essential; what matters is that it remains fixed. Our choice of consistent one-to-one assignment stabilizes the ambiguity, reducing uncertainty in both OOD and task predictions. This contrasts with the design in BackdoorIndicator, which adopts random and dynamic base mappings, leading to greater instability. Instead, our method emphasizes a controlled and interpretable structure, as supported by the empirical results in Figure~\ref{fig:overall_ood_bias_evidence}.


\vspace{0.3em}
\noindent \textbf{Planting a Targeted Watermark Mapping.} Building upon the established base OOD mapping, we further implant a targeted watermark mapping, analogous to the all-to-one setting~\cite{ba_survey_yiming, chameleon} in centralized backdoor:
\begin{align}
    &\pi_2: \mathcal{Y}_o \rightarrow y_{m}, y_{m} \in \mathcal{Y}_t, \quad
    \mathcal{D}_{w} = \{ (\text{tri}(\bm{x}_i, \bm{t}), \pi_2(y_i)) \mid (\bm{x}_i, y_i) \sim \mathcal{D}_p \}, \\
    &\mathcal{L}_{wm} = \sum_{i \in \mathcal{D}_w} \mathrm{CE}(f(\text{tri}(\bm{x}_i, \bm{t}); \bm{w}), \pi_2(y_i)),
\end{align}
Here, $\bm{t}$ denotes the trigger pattern, and $\text{tri}(\cdot, \bm{t})$ represents the trigger implanting function. The watermark mapping function $\pi_2$ assigns all watermark samples to a fixed target label $y_m$. The watermark set $\mathcal{D}_w$ is generated by sampling a fraction $\rho_w$ of the planting set, \ie, $|\mathcal{D}_w| = \rho_w \cdot |\mathcal{D}_p|$. Since the base mapping $\pi_1$ aligns OOD samples with the task feature space, the planted watermark can partially generalize to the task feature distribution. This makes the trigger-label association less tied to specific OOD semantics, thereby enhancing its robustness to OOD prediction bias and reducing the likelihood of being forgotten during benign client training.

Following the prior work~\cite{li2024backdoorindicator}, we save and restore the running mean and variance of BatchNorm layers before and after watermark planting. This preserves the distinct feature distributions of the task and OOD spaces during training. In addition, to prevent the planted watermark from distorting the task-space sample-label mapping, we also apply a global model regularization term. Together, we get the overall loss:
\begin{equation}
    \mathcal{L}_{server} = \mathcal{L}_{base} + \mathcal{L}_{wm} + \lambda\cdot\|\bm{w} - \bm{w}^t\|_2.
\end{equation}

\subsection{Watermark Interaction and Detection}
\noindent \textbf{Watermark Interaction.} After the watermark is injected into the global model $\bm{w}^{t}$, it is delivered to all participating clients for local training. Benign clients perform standard training on their local datasets, optimizing only for task loss. In contrast, malicious clients additionally optimize backdoor poisoning loss. Specifically, we have:
\begin{align}
    &\mathcal{L}_{task} = \sum_{l \in \mathcal{D}_k} \mathrm{CE}(f(\bm{x}_l; \bm{w}^{t+1}_k), y_l), \\
    &\mathcal{L}_{attack} = \mathcal{L}_{task} + \sum_{l \in \mathcal{D}_{k}^{tri}} \mathrm{CE}(f(\text{tri}(\bm{x}_l, \bm{t}); \bm{w}^{t+1}_k), y_a),
\end{align}
where $\mathcal{D}_k$ denotes the $k$-th client's clean data and $\mathcal{D}_{k}^{tri}$ is the triggered subset with label $y_a$ as the attacker's target label.
During training, benign clients only slightly forget the OOD watermark while malicious clients tend to suppress the watermark more aggressively, as their injected backdoors interfere with the watermark.
This divergence in watermark retention behavior serves as the basis for our subsequent backdoor client detection. We provide a detailed analysis of these effects in Supplementary Section~\ref{appendix:collision_ood}.

\vspace{0.3em}
\noindent \textbf{Watermark Detection.} After receiving trained local models from clients, the server evaluates each model's response to the watermark task by measuring the proportion of watermarked samples predicted as the designated label $y_m$. The BatchNorm statistics are also switched from those used for the FL main task to those corresponding to the watermark planting phase during inspection. After collecting the watermark accuracy of all clients, a threshold $\beta$ is applied to distinguish between benign and malicious clients. Thus, clients whose watermark accuracy falls below $\beta$ are excluded from the aggregation process. Thus, we obtain the model aggregation operation as:
\begin{align}
\sum_{k\in \mathcal{S}^{t+1}} \frac{N_k}{N}\cdot \Delta\bm{w}^{t-1}_k \cdot \mathbb{I}[\text{ACC}(f(\mathcal{D}_w;\bm{w}_k^{t+1}), y_{m}) > \beta],
\end{align}

\noindent where $\mathbb{I}$ denotes the indicator function, and ACC represents the watermark accuracy.

\section{Experiments}
\label{sec:experimenal}

\subsection{Main Settings}
\noindent\textbf{Datasets and Models.} 
Image classification has served as a standard benchmark task for evaluating multimedia adversarial and forensic mechanisms. Following common practice in prior work, we conduct all experiments on three image benchmark datasets, EMNIST, CIFAR-10, and CIFAR-100, using the ResNet-18 architecture.

\vspace{0.3em} 
\noindent\textbf{Federated Training Configurations.}
All experiments adopt the standard FedAvg setting \cite{mcmahan2017fedavg}, where we simulate 100 clients in total, with 10 randomly selected to participate in each training round. To simulate the inherent data heterogeneity across clients, we follow \cite{li2020fedprox} to use Dirichlet sampling to partition the dataset. The Dirichlet concentration parameter $\alpha$ controls the label distribution skewness, where smaller values indicate more unbalanced distributions. Specifically, we use $\alpha = 0.9$ by default to simulate mild skewness, and decrease it to 0.3 to model more severe distribution shifts. During local training, each benign client performs 2 epochs with a learning rate of 0.03, while the global training proceeds for 1200 rounds. For conciseness, we omit the less critical parameter details and follow \cite{li2024backdoorindicator} for all settings not explicitly specified.

\vspace{0.3em} 
\noindent\textbf{Backdoor Attack Configurations.}
To ensure a comprehensive evaluation, we consider both single-attacker and multi-attacker scenarios, each covering various attack strategies and backdoor types.
{\textbf{(1)}} In the single-attacker scenario, we adopt a range of attack baselines, including the naive \textit{Vanilla} training \cite{gu2020badnets}, the stealthy \textit{PGD} attack \cite{zitengsun01pgd}, and the latest state-of-the-art methods \textit{Neurotoxin} \cite{Zhengming01Neurotoxin} and \textit{Chameleon} \cite{chameleon}. We adopt three diverse types of backdoor triggers, ranging from the traditional visible trigger (\textit{BadNet} \cite{gu2020badnets}), to the invisible trigger (\textit{Blend} \cite{chen2017targeted}), and the natural feature-based trigger (\textit{Semantic} \cite{bagdasaryan01}), with increasing levels of imperceptibility. {\textbf{(2)}} In the multi-attacker scenario, we increase the ratio of attackers from 30\% to 70\%.
To ensure comprehensive coverage, we include both aligned and divergent attack goals: Uniform \cite{foolsgold} with homogeneous attackers, DBA \cite{xie2019dba} with a shared target but distributed triggers, and NBA \cite{nguyen2024nba} with non-cooperative attackers using distinct backdoors.
Across all settings, the poisoning stage uses a learning rate of 0.03 by default, consistent with benign training, and reduces it by half for more challenging stealthy evaluations.
The backdoor target is class 0 by default. Similar to Indicator, the malicious client launches the attack at the 1000th round, with the poisoning action lasting for 200 rounds.

\vspace{0.3em} 
\noindent\textbf{Backdoor Detection Configurations.}
We evaluate six representative federated backdoor detection methods that span a range of strategies, from classical model-space analysis to more advanced and proactive techniques: We begin with traditional methods that rely on the statistical properties of model updates:
\textbf{(1)} \textit{MultiKrum} \cite{multiKrum} detects backdoor models based on the Euclidean distance among client updates.
\textbf{(2)} \textit{Foolsgold} \cite{foolsgold} identifies malicious clients by analyzing pairwise cosine similarity to detect overly aligned update directions.
\textbf{(3)} \textit{Rflbat} \cite{wang2022rflbat} projects model updates into a PCA space and flags outliers as potential attackers. Then we further include more advanced approaches that incorporate prediction behavior or mitigation mechanisms:
\textbf{(4)} \textit{Deepsight} \cite{rieger2022deepsight} evaluates similarity in model predictions and output neuron activations to reveal abnormal client behavior.
\textbf{(5)} \textit{Flame} \cite{nguyen2022flame} combines detection based on model similarity together with mitigation techniques such as gradient clipping and noise injection. Finally, we include the pioneering proactive method that shifts the detection paradigm:
\textbf{(6)} \textit{BackdoorIndicator} \cite{li2024backdoorindicator} injects OOD patterns into the global model and identifies backdoor models based on their reactions to these patterns.

\vspace{0.3em} 
\noindent\textbf{Configuration of Coward.}
Here we detail the key factors related to Coward. To construct the OOD planting set, we use EMNIST (expanded to three channels) when the main task is CIFAR-10 or CIFAR-100, and grayscale CIFAR-10 when the main task is EMNIST.
The same planting set is used for Indicator for fairness.
By default, the size of the planting set is 1000.
For watermark injection, we use 20\% of the planting set samples for trigger mapping, with a learning rate of 0.001 for 5 iterations.
In our main experiments, we adopt WaNet \cite{nguyen2021wanet} as the default watermark trigger.
For detection, a fixed 5\% threshold is used across all settings: clients with watermark accuracy below this are identified as malicious. The watermark target is 8 by default.

\vspace{0.3em} 
\noindent\textbf{Evaluation Metrics.}
We evaluate the detection performance using true positive rate (TPR) and false positive rate (FPR). TPR reflects the proportion of correctly identified malicious clients, while FPR represents the proportion of benign clients incorrectly classified as malicious. Higher TPR and lower FPR indicate more effective detection. In addition, we also assess attack success rate (ASR) and benign accuracy (BA) after detection, where lower ASR and higher BA indicate that the model's primary functionality is preserved, and backdoor effects are effectively excluded from the model.

\subsection{Main Results}
\label{sec:mainresults}
We comprehensively evaluate all baseline detection methods from three aspects:

\vspace{0.3em}
\noindent \textbf{Resistance to FL Non-i.i.d.}
We evaluate all baseline methods under varying degrees of non-i.i.d. data across three datasets, using a vanilla backdoor setting with a fixed spot-pattern trigger and a learning rate matching that of benign clients. As shown in Table~\ref{tab:noniid}, \texttt{Coward} consistently achieves the best overall performance across all levels of heterogeneity, with TPRs above 95\% and FPRs below 10.5\% in all cases. Besides the overall conclusion, we draw several key observations: \textbf{(1)} \emph{\texttt{Coward} significantly reduces FPR across all settings compared to BackdoorIndicator}, highlighting its advantage in mitigating OOD bias. A more detailed analysis is provided in Section~\ref{subsec:effectiveness}. \textbf{(2)} \emph{Proactive methods significantly outperform passive ones under severe data heterogeneity}. Taking the most challenging non-i.i.d. setting ($\alpha=0.3$) as an example, aside from our method, BackdoorIndicator also outperforms all passive baselines, achieving TPRs of 93\% on CIFAR-10 and 96\% on EMNIST. This result supports the robustness of proactive detection in heterogeneous environments. In contrast, passive methods exhibit significant performance degradation. For instance, the state-of-the-art passive method Flame sees its TPR drop from above 95\% to below 10\% on both CIFAR-100 and EMNIST, which aligns with our earlier analysis. \textbf{(3)} \textit{Some passive methods fail due to their unrealistic assumptions}. For example, Rflbat performs poorly across all heterogeneity levels, likely because it relies only on the top two principal components of model parameters, losing critical information, especially under our challenging FL setup. Similarly, FoolsGold collapses on CIFAR-100, always judging all clients as benign (resulting in 0\% TPR and 0\% FPR). This failure stems from two factors: the high-dimensional 100-class prediction vector and its reliance on multi-attacker scenarios. These conditions lead to extremely low cosine similarity values, causing an all-as-benign judgment which is unhelpful for detecting malicious clients.

\begin{table}[!t]
\caption{Defense results against vanilla single-client backdoor attack under different non-i.i.d. settings. \textbf{Boldface} indicates the best result value (excluding the collapsed case with TPR/FPR=0/0), while \red{red} highlights metric-specific failures based on a 50\% threshold (\ie, values above 50\% for FPR and ASR, and below 50\% for TPR and BA).}
\label{tab:noniid}
\centering
\scriptsize
\text{(a) Detection performance.}
\vspace{0.2em}

\setlength{\tabcolsep}{0.8pt}
\begin{adjustbox}{width=\linewidth}
\begin{tabular}{cc|*{14}{c}}
\toprule
\multirow{2}{*}{\text{Dataset}} & \multirow{2}{*}{\text{NonIID}}
& \multicolumn{2}{c}{\text{MultiKrum}}
& \multicolumn{2}{c}{\text{FoolsGold}}
& \multicolumn{2}{c}{\text{Rflbat}}
& \multicolumn{2}{c}{\text{DeepSight}}
& \multicolumn{2}{c}{\text{Flame}}
& \multicolumn{2}{c}{\text{Indicator}}
& \multicolumn{2}{c}{\textbf{Ours}} \\
\cmidrule(lr){3-4}\cmidrule(lr){5-6}\cmidrule(lr){7-8}\cmidrule(lr){9-10}\cmidrule(lr){11-12}\cmidrule(lr){13-14}\cmidrule(lr){15-16}
& & TPR & FPR
& TPR & FPR
& TPR & FPR
& TPR & FPR
& TPR & FPR
& TPR & FPR
& TPR & FPR
\\
\midrule
\multirow{3}{*}{\text{CIFAR10}} & \text{0.9}
& 93.5 & 34.1
& \red{44.0} & {44.3}
& \red{2.5} & 9.6
& 74.0 & \red{60.3}
& \textbf{99.0} & 33.4
& 97.5 & 23.0
& \textbf{99.0} & \textbf{3.9} \\
& \text{0.6}
& 62.0 & 37.6
& \red{47.5} & \red{50.3}
& \red{1.5} & 10.4
& 66.0 & \red{53.2}
& 69.0 & 36.8
& 94.0 & 26.1
& \textbf{99.0} & \textbf{9.2} \\
& \text{0.3}
& {57.5} & 38.1
& 60.5 & \red{57.2}
& \red{1.0} & 9.2
& 57.0 & {44.8}
& 62.0 & 37.5
& 93.0 & 31.0
& \textbf{99.0} & \textbf{9.6} \\
\midrule
\multirow{3}{*}{\text{CIFAR100}} & \text{0.9}
& \textbf{98.0} & 22.4
& \red{0.0} & 0.0
& \red{12.0} & 22.6
& \red{39.0} & 32.6
& 97.0 & 22.6
& 95.0 & 47.0
& \textbf{98.0} & \textbf{2.9} \\
& \text{0.6}
& 96.0 & 22.7
& \red{0.0} & 0.0
& \red{11.0} & 22.4
& 55.0 & 47.8
& 95.0 & 22.8
& 81.0 & 47.8
& \textbf{100.0} & \textbf{4.3} \\
& \text{0.3}
& \red{2.0} & 33.1
& \red{0.0} & 0.0
& \red{6.0} & 23.3
& 75.0 & \red{65.3}
& \red{5.0} & {32.8}
& 76.0 & 43.6
& \textbf{98.0} & \textbf{8.6} \\
\midrule
\multirow{3}{*}{\text{EMNIST}} & \text{0.9}
& \textbf{100.0} & 22.2
& {52.0} & 39.1
& \red{16.0} & 22.7
& {57.0} & {45.9}
& \textbf{100.0} & 22.0
& 99.0 & \red{57.8}
& 98.0 & \textbf{1.6} \\
& \text{0.6}
& 98.0 & 22.4
& \red{35.0} & {41.7}
& \red{11.0} & 23.1
& {56.0} & {49.4}
& 97.0 & 22.3
& {96.0} & \red{52.7}
& \textbf{98.0} & \textbf{1.6} \\
& \text{0.3}
& \red{9.0} & 32.3
& \red{40.0} & {48.2}
& \red{1.0} & 25.1
& \red{49.0} & {47.0}
& \red{9.0} & {32.1}
& 96.0 & {45.6}
& \textbf{98.0} & \textbf{8.3} \\
\bottomrule
\end{tabular}

\end{adjustbox}

\vspace{0.6em}
\text{(b) Attack success rate and benign accuracy.}
\vspace{0.2em}

\setlength{\tabcolsep}{0.8pt}
\begin{adjustbox}{width=\textwidth}
\begin{tabular}{cc|*{14}{c}}
\toprule
\multirow{2}{*}{\text{Dataset}} & \multirow{2}{*}{\text{NonIID}} & \multicolumn{2}{c}{\text{MultiKrum}} & \multicolumn{2}{c}{\text{FoolsGold}} & \multicolumn{2}{c}{\text{Rflbat}} & \multicolumn{2}{c}{\text{DeepSight}} & \multicolumn{2}{c}{\text{Flame}} & \multicolumn{2}{c}{\text{Indicator}} & \multicolumn{2}{c}{\textbf{Ours}} \\
\cmidrule(lr){3-4}\cmidrule(lr){5-6}\cmidrule(lr){7-8}\cmidrule(lr){9-10}\cmidrule(lr){11-12}\cmidrule(lr){13-14}\cmidrule(lr){15-16}
  &  & ASR & BA
 & ASR & BA
 & ASR & BA
 & ASR & BA
 & ASR & BA
 & ASR & BA
 & ASR & BA
 \\
\midrule
\multirow{3}{*}{\text{CIFAR10}} & \text{0.9} & 34.1 & \textbf{92.1} & \red{91.3} & \textbf{{92.1}} & \red{91.3} & {91.8} & \red{58.2} & 91.9 & 9.9 & \textbf{92.1} & \textbf{9.4} & 91.2 & 10.3 & 90.5 \\
 & \text{0.6} & \red{88.9} & 91.9 & \red{89.5} & {91.4} & \red{91.3} & {\textbf{92.0}} & \red{66.4} & \textbf{92.0} & \red{87.2} & 91.8 & 35.3 & 91.5 & \textbf{9.9} & 90.8 \\
 & \text{0.3} & \red{90.2} & 91.7 & \red{84.5} & 90.8 & \red{91.2} & {\textbf{91.9}} & \red{72.7} & {91.8} & \red{88.5} & 91.8 & 24.2 & 91.5 & \textbf{10.0} & 90.4 \\
\midrule
\multirow{3}{*}{\text{CIFAR100}} & \text{0.9} & 1.1 & 69.7 & \red{82.9} & {\textbf{69.9}} & \red{83.2} & 69.7 & {46.8} & 69.8 & 1.1 & 69.7 & 1.3 & 69.0 & \textbf{1.0} & 69.1 \\
 & \text{0.6} & \textbf{1.0} & \textbf{69.9} & \red{82.7} & {69.5} & \red{83.4} & {69.6} & {41.5} & {69.5} & \textbf{1.0} & 69.8 & \red{56.1} & 69.0 & \textbf{1.0} & 69.2 \\
 & \text{0.3} & \red{82.0} & {69.5} & \red{81.5} & {69.3} & \red{82.4} & {69.2} & {25.2} & \textbf{69.7} & \red{81.6} & {69.3} & 46.8 & 67.8 & \textbf{1.0} & 68.9 \\
\midrule
\multirow{3}{*}{\text{EMNIST}} & \text{0.9} & \textbf{10.0} & \textbf{99.8} & \red{99.9} & {99.7} & \red{100.0} & {99.7} & {27.4} & {99.7} & \textbf{10.0} & 99.7 & \textbf{10.0} & 99.6 & \textbf{10.0} & 99.7 \\
 & \text{0.6} & \textbf{10.0} & \textbf{99.8} & \red{99.7} & {99.7} & \red{99.8} & {99.7} & {47.8} & {99.7} & \textbf{10.0} & 99.7 & {10.8} & 99.6 & \textbf{10.0} & 99.7 \\
 & \text{0.3} & \red{100.0} & {99.7} & \red{99.5} & {\textbf{99.8}} & \red{99.9} & {99.7} & {47.3} & {\textbf{99.8}} & \red{100.0} & {99.7} & {10.2} & 99.6 & \textbf{10.0} & 99.7 \\
\bottomrule
\end{tabular}
\end{adjustbox}
\vspace{-1.5em}
\end{table}

\vspace{0.3em}
\noindent \textbf{Resistance to Advanced Attacks}.
We evaluate all baselines under more challenging, advanced, and stealthy attack settings.
Specifically, the malicious client adopts a more imperceptible noise-blended trigger and reduced local learning rate in all attack methods.
We report detection performance in Table~\ref{tab:mainexp}, while additional ASR and BA metrics are reported in Supplementary Table~\ref{tab:mainexp_asr_ba}, as they exhibit trends consistent with those in Table~\ref{tab:noniid}.
As the results show, {\textit{our method achieves the best overall performance, with proactive detection showing strong robustness, while passive methods suffer substantial degradation}}.
This highlights the advantage of the proactive paradigm in handling sophisticated attacks.
\textbf{(1)} \textit{Among proactive methods}, \texttt{Coward} consistently achieves the best performance, maintaining close to 100\% TPR and sub-10\% FPR across all advanced attack types, including state-of-the-art strategies such as Neurotoxin and Chameleon.
BackdoorIndicator also performs strongly, with TPRs consistently exceeding 85\% in most settings; however, it suffers from relatively high FPRs.
\textbf{(2)} \textit{For passive methods}, performance degradation is anticipated under such stealthy attacks, since the attacker introduces only subtle changes to the model, which reduces its suspiciousness.
As an instance, \textit{MultiKrum}, which selects representative updates based on Euclidean distance, becomes highly ineffective, with TPRs falling below 12.5\% across all stealthy attack settings.
\textit{Flame}, which filters out updates with large angular deviations, also suffers notable degradation, with TPRs dropping below 30.5\% on the CIFAR-10 dataset.
In contrast, \textit{DeepSight} shows better resilience, which performs consistently with TPRs over 40\% in most attack cases.
This probably stems from its joint use of both parameter and prediction discrepancies.

\begin{table}[!t]
\caption{Detection performance against stealthy single-client backdoor attacks with advanced training strategies. Result highlighting (in \textbf{boldface} and in \red{red}) follows the same rules as those described in Table \ref{tab:noniid}. NTX and CML stand for Neurotoxin and Chameleon, respectively.}
\label{tab:mainexp}
\centering
\scriptsize
\setlength{\tabcolsep}{1pt}
\vspace{0.2em}

\begin{adjustbox}{width=\textwidth}
\begin{tabular}{cc|*{14}{c}}
\toprule
\multirow{2}{*}{\text{Dataset}} & \multirow{2}{*}{\text{Attack}}
& \multicolumn{2}{c}{\text{MultiKrum}}
& \multicolumn{2}{c}{\text{FoolsGold}}
& \multicolumn{2}{c}{\text{Rflbat}}
& \multicolumn{2}{c}{\text{DeepSight}}
& \multicolumn{2}{c}{\text{Flame}}
& \multicolumn{2}{c}{\text{Indicator}}
& \multicolumn{2}{c}{\textbf{Ours}} \\
\cmidrule(lr){3-4}\cmidrule(lr){5-6}\cmidrule(lr){7-8}\cmidrule(lr){9-10}\cmidrule(lr){11-12}\cmidrule(lr){13-14}\cmidrule(lr){15-16}
& & TPR & FPR
& TPR & FPR
& TPR & FPR
& TPR & FPR
& TPR & FPR
& TPR & FPR
& TPR & FPR
\\
\midrule
\multirow{4}{*}{\text{CIFAR10}} & \text{Vanilla}
& \textcolor{red}{5.5} & 32.7
& 53.5 & 44.4
& \red{3.0} & 24.2
& 77.9 & \red{64.0}
& \textcolor{red}{30.5} & 29.9
& 93.0 & 34.7
& \textbf{100.0} & \textbf{3.3} \\
& \text{PGD}
& \textcolor{red}{5.5} & 32.7
& 53.5 & 44.6
& \textcolor{red}{2.0} & 24.5
& 75.5 & \red{61.0}
& \textcolor{red}{12.0} & 31.8
& 92.0 & 34.1
& \textbf{97.5} & \textbf{4.2} \\
& \text{NTX}
& \textcolor{red}{6.0} & 32.7
& \textcolor{red}{15.5} & 43.1
& \textcolor{red}{2.5} & 24.3
& 69.2 & \textcolor{red}{58.5}
& \textcolor{red}{30.0} & 29.9
& 52.5 & 31.7
& \textbf{99.5} & \textbf{7.8} \\
& \text{CML}
& \textcolor{red}{8.0} & 32.4
& \textcolor{red}{46.5} & 44.1
& \textcolor{red}{3.5} & 23.6
& \textcolor{red}{39.0} & 36.5
& \textcolor{red}{10.0} & 32.2
& 87.0 & 33.8
& \textbf{94.5} & \textbf{9.1} \\
\midrule
\multirow{4}{*}{\text{CIFAR100}} & \text{Vanilla}
& \textcolor{red}{6.5} & 32.6
& \textcolor{red}{0.0} & {0.0}
& \textcolor{red}{3.5} & 23.7
& \textcolor{red}{48.7} & 44.4
& \textcolor{red}{7.5} & 32.5
& 91.0 & 45.3
& \textbf{100.0} & \textbf{1.9} \\
& \text{PGD}
& \textcolor{red}{6.5} & 32.6
& \textcolor{red}{0.0} & {0.0}
& \textcolor{red}{2.5} & 23.4
& \textcolor{red}{46.7} & 43.0
& \textcolor{red}{2.0} & 33.1
& 88.0 & 45.7
& \textbf{100.0} & \textbf{5.8} \\
& \text{NTX}
& \textcolor{red}{7.0} & 32.6
& \textcolor{red}{0.0} & {0.0}
& \textcolor{red}{4.5} & 23.2
& 51.3 & 42.7
& \textcolor{red}{2.5} & 33.1
& 96.0 & 46.9
& \textbf{100.0} & \textbf{9.4} \\
& \text{CML}
& \red{0.0} & 33.3
& \red{0.0} & {0.0}
& \red{5.0} & 24.3
& \red{41.0} & 34.7
& \red{0.0} & 33.3
& \textbf{98.5} & 47.5
& 98.2 & \textbf{7.3} \\
\midrule
\multirow{4}{*}{\text{EMNIST}} & \text{Vanilla}
& \textcolor{red}{8.5} & 32.4
& \textcolor{red}{12.0} & 35.7
& \textcolor{red}{1.0} & 24.6
& \textcolor{red}{24.0} & 37.3
& \textcolor{red}{44.5} & 28.3
& 88.3 & {45.7}
& \textbf{100.0} & \textbf{3.6} \\
& \text{PGD}
& \textcolor{red}{10.0} & 32.2
& \textcolor{red}{11.0} & 35.2
& \textcolor{red}{2.0} & 11.7
& \textcolor{red}{44.0} & 46.9
& \textcolor{red}{37.0} & 31.4
& 88.5 & {46.0}
& \textbf{100.0} & \textbf{4.3} \\
& \text{NTX}
& \textcolor{red}{10.5} & 31.9
& \textcolor{red}{2.5} & 36.3
& \textcolor{red}{1.5} & 12.0
& \textcolor{red}{38.5} & 43.1
& \textcolor{red}{44.5} & 17.2
& 88.6 & 45.3
& \textbf{99.0} & \textbf{3.6} \\
& \text{CML}
& \textcolor{red}{12.5} & 31.9
& \textcolor{red}{21.5} & 37.2
& \textcolor{red}{3.0} & 24.8
& \textcolor{red}{37.5} & 42.7
& 98.0 & 22.3
& 91.5 & \red{51.7}
& \textbf{99.5} & \textbf{2.4} \\
\bottomrule
\end{tabular}
\end{adjustbox}
\vspace{-1.5em}

\end{table}

\vspace{0.3em}
\noindent \textbf{Robustness to Multiple Attackers}.
Table~\ref{tab:multi} presents detection performance under multi-attacker scenarios.
Additional ASR and BA metrics are reported in Supplementary Table~\ref{tab:multi_asr_ba}, as they exhibit trends consistent with those in Table~\ref{tab:noniid}.
As shown in this table, proactive methods exhibit better overall performance across all three settings.
This is expected as \textit{the detection mechanism of proactive approaches is independent of the number of attackers}.
Among them, our method achieves the best overall performance, consistently maintaining TPRs above 99\% and FPRs below 15\%.
BackdoorIndicator also performs well but becomes less effective when the attacker proportion reaches 70\%.
In such cases, a temporary detection failure allows the increased number of attackers to rapidly reinforce the global backdoor, resulting in a higher ASR.
Moreover, the retained backdoor could be further maintained by the indicator’s maintenance effect.
Among passive baselines, FoolsGold performs best under Uniform and DBA attacks, achieving a TPR of 100\%, benefiting from its assumption of highly similar, coordinated malicious updates.
However, it fails in uncoordinated scenarios, where attackers pursue different targets.
In such cases, the TPR drops below 40\% as the method's core assumption of shared attack objectives no longer holds.
Outlier-based methods such as MultiKrum and Flame become ineffective when attackers control 70\% of clients, with TPRs dropping under 42\% in all cases.
These methods assume malicious updates are outliers, which fails when malicious clients form the majority.

\begin{table}[!t]
\caption{Detection performance under multi-client backdoor attacks on the CIFAR10 dataset. Result highlighting (in \textbf{boldface} and in \red{red}) follows the same rules in Table \ref{tab:noniid}.}
\label{tab:multi}
\centering
\scriptsize
\setlength{\tabcolsep}{1pt}
\vspace{0.2em}

\begin{adjustbox}{width=\textwidth}
\begin{tabular}{cc|*{14}{c}}
\toprule
\multirow{2}{*}{\text{Strategy}} & \multirow{2}{*}{\text{Atk\%}}
& \multicolumn{2}{c}{\text{MultiKrum}}
& \multicolumn{2}{c}{\text{FoolsGold}}
& \multicolumn{2}{c}{\text{Rflbat}}
& \multicolumn{2}{c}{\text{DeepSight}}
& \multicolumn{2}{c}{\text{Flame}}
& \multicolumn{2}{c}{\text{Indicator}}
& \multicolumn{2}{c}{\textbf{Ours}} \\
\cmidrule(lr){3-4}\cmidrule(lr){5-6}\cmidrule(lr){7-8}\cmidrule(lr){9-10}\cmidrule(lr){11-12}\cmidrule(lr){13-14}\cmidrule(lr){15-16}
& & TPR & FPR
& TPR & FPR
& TPR & FPR
& TPR & FPR
& TPR & FPR
& TPR & FPR
& TPR & FPR
\\
\midrule
\multirow{3}{*}{\text{Uniform}} & \text{30\%}
& \textbf{100.0} & \red{71.4}
& \textbf{100.0} & 34.2
& \textcolor{red}{39.4} & 22.5
& \textcolor{red}{3.7} & 14.6
& \textbf{100.0} & 14.3
& 99.2 & 29.5
& \textbf{100.0} & \textbf{6.7} \\
& \text{50\%}
& \textcolor{red}{3.0} & \red{97.0}
& \textbf{100.0} & 27.1
& \red{31.3} & \textbf{0.0}
& \red{1.1} & 24.0
& \red{21.5} & \red{57.3}
& 99.1 & 16.3
& \textbf{100.0} & 8.2 \\
& \text{70\%}
& \red{29.0} & \red{98.9}
& \textbf{100.0} & 12.8
& \red{14.3} & 11.8
& \red{2.4} & \red{54.8}
& \red{31.3} & \red{57.8}
& 99.4 & 44.6
& \textbf{100.0} & \textbf{11.2} \\
\midrule
\multirow{3}{*}{\text{DBA}} & \text{30\%}
& \textbf{100.0} & \textcolor{red}{57.4}
& \textbf{100.0} & 32.9
& \textcolor{red}{28.0} & 15.4
& \textcolor{red}{5.6} & 12.4
& \textbf{100.0} & 14.3
& 98.2 & 18.4
& \textbf{100.0} & \textbf{9.7} \\
& \text{50\%}
& \red{22.8} & \textcolor{red}{77.2}
& \textbf{100.0} & 26.7
& \red{29.2} & 0.0
& \red{2.9} & 20.1
& 60.5 & 18.9
& 98.6 & 16.0
& \textbf{100.0} & \textbf{9.9} \\
& \text{70\%}
& \red{31.1} & \red{94.1}
& \textbf{100.0} & 12.8
& \red{9.5} & 2.8
& \red{3.4} & {44.7}
& \red{42.0} & 34.8
& 98.8 & 15.8
& \textbf{100.0} & \textbf{11.2} \\
\midrule
\multirow{3}{*}{\text{NBA}} & \text{30\%}
& \textbf{100.0} & \red{57.1}
& \red{37.5} & {48.4}
& {78.5} & \textcolor{red}{70.8}
& \red{2.8} & 31.2
& 55.2 & 18.9
& 97.8 & 39.4
& {98.7} & \textbf{7.1} \\
& \text{50\%}
& 63.7 & 36.3
& \textcolor{red}{33.3} & \textcolor{red}{54.6}
& \textcolor{red}{16.5} & 15.8
& \textcolor{red}{1.4} & 48.1
& \textcolor{red}{19.2} & 40.6
& 93.7 & 38.2
& \textbf{98.5} & \textbf{15.0} \\
& \text{70\%}
& \textcolor{red}{36.9} & \textcolor{red}{80.7}
& \textcolor{red}{23.4} & \textcolor{red}{62.7}
& \textcolor{red}{0.1} & \textbf{0.0}
& \textcolor{red}{1.3} & \textcolor{red}{78.8}
& \textcolor{red}{10.6} & \textcolor{red}{74.8}
& 81.3 & 14.0
& \textbf{86.2} & 14.6 \\
\bottomrule
\end{tabular}
\end{adjustbox}
\vspace{-1.5em}
\end{table}

\begin{table}[!t]
\caption{Performance on different OOD planting sets. Result highlighting (in \textbf{boldface} and in \red{red}) follows the same rules in Table \ref{tab:noniid}.
}
\vspace{0.2em}
\label{tab:oodset}
\centering
\setlength{\tabcolsep}{1.5pt}

\begin{adjustbox}{max width=0.8\textwidth}
\begin{tabular}{cc|ccccc|cccc}
\toprule
\multirow{2}{*}{{Main Task}} &
\multirow{2}{*}{\text{Planting Set}} & \multicolumn{4}{c}{\text{Indicator}}  && \multicolumn{4}{c}{\textbf{Ours}}\\

\cline{3-6} \cline{8-11} & &
TPR$\uparrow$ & FPR$\downarrow$ & ASR$\downarrow$ & BA$\uparrow$ &&
TPR$\uparrow$ & FPR$\downarrow$ & ASR$\downarrow$ & BA$\uparrow$
\\ \hline

\multirow{3}{*}{\text{CIFAR10}}  &
\text{EMNIST} & 98.0 & 24.0 & 10.2 & 91.6  &&
\textbf{99.5} & \textbf{7.3} & \textbf{9.7} & \textbf{91.8} \\

& \text{CIFAR100} &
97.0 & 14.2 & 10.0 & \textbf{91.7} &&
\textbf{99.5} & \textbf{6.7} & \textbf{9.9} & 91.2  \\

& \text{NOISE} &
\textbf{100.0} & \red{57.4} & \textbf{9.8} & \textbf{90.7} &&
\textbf{100.0} & \textbf{12.7} & 10.1 & 90.6 \\ \hline

\multirow{3}{*}{\text{CIFAR100}} & \text{EMNIST} &
98.0 & 34.0 & 1.3 & 68.2 &&
\textbf{100.0} & \textbf{5.3} & \textbf{1.0} & \textbf{68.9} \\

& \text{CIFAR10}&
99.0 & 44.6 & 1.8 & 67.2 &&
\textbf{100.0} & \textbf{6.3} & \textbf{1.2} & \textbf{67.5}  \\

& \text{NOISE} &
98.0 & \red{60.0} & 1.5  & \textbf{67.9} &&
\textbf{100.0} & \textbf{4.4} & \textbf{1.0} & 67.8 \\ \hline

\multirow{3}{*}{\text{EMNIST}} & \text{CIFAR10}  &
95.0 & \red{57.8} & \textbf{10.0} & 99.6  &&
\textbf{99.5} & \textbf{2.2} & \textbf{10.0} & \textbf{99.8} \\

& \text{CIFAR100} &
97.0 & \red{58.7} & \textbf{10.0} & 99.6 &&
\textbf{99.0} & \textbf{1.5} & 10.1 & \textbf{99.7} \\

& \text{NOISE} &
100.0 & \red{73.5} & \textbf{10.0} & \textbf{99.6} &&
\textbf{99.0} & \textbf{4.8} & 10.2 & \textbf{99.6} \\

\bottomrule

\end{tabular}
\end{adjustbox}
\vspace{-1.5em}
\end{table}

\subsection{Ablation Study}
We investigate key factors that influence watermark planting and detection: \textbf{(1)} the choice of OOD planting set, \textbf{(2)} the pairing of watermark and backdoor triggers, and \textbf{(3)} the selection of detection thresholds. For a more comprehensive analysis, please refer to Supplementary Sections~\ref{appendix:wm_cfg} and~\ref{appendix:wm_inj}.

\vspace{0.3em}
\noindent \textbf{Robustness to Choices of OOD Datasets}. Table~\ref{tab:oodset} presents detection performance across diverse OOD planting sets, including task-similar data, distinctly different data, and an extreme case with synthetic Gaussian noise.
We observe that \texttt{Coward} consistently maintains low FPR across all settings, while Indicator's performance degrades as the OOD dataset diverges from the main task.
Using the most challenging main task, the CIFAR-100 dataset, as an example, \texttt{Coward}'s FPR remains below 6.3\% across all planting sets.
In contrast, Indicator's FPR increases sharply from 44.6\% on CIFAR-10 to 60\% with Gaussian noise.
This confirms our earlier analysis of Indicator's increased vulnerability to OOD bias.
Overall, these results show that \texttt{Coward} is more robust to OOD data selection, offering greater flexibility in real-world deployment.

\vspace{0.3em}
\noindent \textbf{Robustness to Choices of Triggers}.  
Table~\ref{tab:triggerpair} summarizes performance under various combinations of trigger types.
The attacker's trigger varies in increasing stealthiness: from a visible pixel pattern, to an invisible blend pattern, and finally to an implicit semantic pattern.
The watermark trigger also spans a range of types, including four visible patterns (diagonal, square, triangular, and noise), as well as two invisible variants: a mosaic pattern and the sample-specific WaNet pattern.
We see that \texttt{Coward} performs consistently well across all settings, achieving a TPR above 96.5\% and an FPR below 17.3\%.
These results demonstrate our method’s robustness to diverse triggers and the generalizability of the multi-backdoor collision effect.

\begin{table}[t]
\caption{Detection performance with different combinations of watermark trigger \textbf{(S)} and attackers' client-side malicious trigger \textbf{(A)} on the CIFAR-10 dataset.
}
\vspace{0.2em}
\label{tab:triggerpair}
\centering
\setlength{\tabcolsep}{1pt}
\renewcommand{\arraystretch}{0.8}
\begin{adjustbox}{max width=0.9\textwidth}

\begin{tabular}{c|cccc|cccc|cccc}
\toprule
\multirow{2}{*}{\textbf{$\downarrow$\textbf{S} / \textbf{A}$\rightarrow$}} & \multicolumn{4}{c}{\textbf{Pixel} (visible)} &
\multicolumn{4}{c}{\textbf{Blend} (invisible)} &
\multicolumn{4}{c}{\textbf{Semantic} (implicit)} \\

\cline{2-5} \cline{6-9} \cline{10-13}
& TPR$\uparrow$ & FPR$\downarrow$ & ASR$\downarrow$ & BA$\uparrow$ &
TPR$\uparrow$ & FPR$\downarrow$ & ASR$\downarrow$ & BA$\uparrow$ &
TPR$\uparrow$ & FPR$\downarrow$ & ASR$\downarrow$ & BA$\uparrow$ \\
\midrule
\text{Diagonal}  & 100.0 & 6.1 & 9.8 & 91.6 &
99.5 & 6.6 & 5.2 & 91.7 &
98.5 & 17.3 & 0.0 & 91.8 \\
\text{Square}    & 100.0 & 9.4 & 9.6 & 91.4 &
99.5 & 9.1 & 4.8 & 91.4 &
98.5 & 16.2 & 0.0 & 91.4 \\
\text{Triangle}  & 99.5 & 4.6 & 9.8  & 91.6 &
98.5 & 6.2 & 7.8 & 91.7 &
99.0 & 12.3 & 0.0 & 90.8 \\
\text{Noise}     & 99.5 & 6.1 & 9.9  & 91.5 &
99.0 & 6.7 & 7.9 & 91.5 &
99.5 & 7.3 & 0.0  & 91.4 \\
\text{Mosaic}    & 96.5 & 10.6 & 18.0 & 91.6 &
99.5 & 12.0 & 4.6 & 91.3 &
99.0 & 14.2 & 0.0 & 90.7 \\
\text{WaNet}     & 99.5 & 7.3 & 9.8  & 91.4 &
99.5 & 6.8 & 5.4 & 91.5 &
98.0 & 11.9 & 0.0 & 90.8 \\

\bottomrule
\end{tabular}
\end{adjustbox}
\vspace{-1.5em}
\end{table}

\begin{table}[t]
  \centering
  \caption{Performance of \texttt{Coward} under different watermark-retention thresholds on the CIFAR-10 dataset.}
  \vspace{0.2em}
  \label{tab:wm_thresh}
  \renewcommand{\arraystretch}{0.8}
  \setlength{\tabcolsep}{1pt}
  \begin{adjustbox}{max width=0.5\textwidth}
  \begin{tabular}{c|cccc}
      \toprule
      \text{Threshold} & \text{TPR}$\uparrow$ & \text{FPR}$\downarrow$ & \text{ASR}$\downarrow$ & \text{BA}$\uparrow$ \\
      \midrule
      1\% & 98.0 & 1.3 & 10.2 & 91.5 \\
      5\% & 100.0 & 2.6 & 10.1 & 91.5 \\
      10\%& 100.0 & 8.6 & 10.1 & 91.4 \\
      15\%& 100.0 & 12.0 & 9.9 & 91.5 \\
      20\%& 100.0 & 13.9 & 10.0 & 91.2 \\
      \bottomrule
    \end{tabular}
  \end{adjustbox}
  \vspace{-1.5em}
\end{table}

\vspace{0.3em}
\noindent \textbf{Robustness to Choices of Detection Thresholds}. 
We further evaluate the impact of the detection threshold on the effectiveness of \texttt{Coward}. Since both attackers and benign clients may disrupt the watermark to some extent, the optimal threshold tends to fall within a relatively low range. 
Thus, we conduct threshold analysis within a mild range from 1\% to 20\%. As shown in Table~\ref{tab:wm_thresh}, varying the detection threshold has minimal impact on TPR, which consistently remains above 98\%. However, it progressively increases the FPR, rising from 1.0\% to 13.9\%. This increase is primarily due to client-side data heterogeneity. Certain benign clients, particularly those with severely skewed local distributions, may forget the watermark and exhibit significantly lower watermark accuracy than the majority. In contrast, malicious clients consistently show very low watermark accuracy. Empirically, we find that setting the threshold at 5\% offers a reliable and practical guideline for a balanced TPR and FPR across diverse scenarios.

\subsection{Resistance to Potential Adaptive Attack}

\begin{table}[t]
\caption{Performance of \texttt{Coward} under the adaptive attack (AA). The subscripts $l$ and $g$ denote the ASR of the attacker's backdoor evaluated on the malicious local model and the aggregated global model, respectively.}
\label{tab:adaptive}
\vspace{0.2em}

\centering
\setlength{\tabcolsep}{2pt}
\renewcommand{\arraystretch}{0.8}
\begin{adjustbox}{max width=0.7\textwidth}

\begin{tabular}{c|cc|cc}
\toprule
\text{Method} & \text{TPR}$\uparrow$  & \text{FPR}$\downarrow$ & \text{ASR}$_g \downarrow$ & \text{ASR}$_l$ \\
\hline
w./o. defense & /    & /   & 91.2 & 95.6 \\
w./o. AA.     & 100.0  & 5.0  & 10.1 & 98.9 \\
\hline
AA. ($T=5$) & 99.5 & 4.4 & 10.1 & 90.4 \\
AA. ($T=1$) & 62.5 & 5.4 & 22.9 & 73.9 \\
\bottomrule
\end{tabular}
\end{adjustbox}
\vspace{-1.5em}
\end{table}

\noindent \textbf{Potential Adaptive Attack Strategy.} We consider adaptive attackers who attempt to bypass our collision-based detection by injecting a similar watermark to preserve high watermark accuracy. Since the watermark target is chosen from a finite, known set, the attacker has a non-trivial chance of guessing it. To model this, we design a {\textit{periodic guessing attack}} with switching period $T$, where the attacker updates the guessed label every $T$ rounds. Smaller $T$ increases the chance of early success but may cause unstable feedback, while larger $T$ offers more stable signals at the cost of later success.

\vspace{0.3em}
\noindent \textbf{Results.} As shown in Table~\ref{tab:adaptive}, the guessing attack appears to reduce the TPR but does not truly bypass the detection. For instance, when $T = 1$, the TPR drops from 100\% to 62.5\%, indicating some disruption due to a correct guess. However, the attacker's main objective is to increase the ASR on the aggregated global model. Notably, the global ASR remains low across all adaptive settings, reaching only 22.9\% at most when $T = 1$, which falls short of being successful. Upon further analysis, we find that the guessing attack inherently suffers from a fundamental contradiction, which we refer to as the \textit{{local collision contradiction}}. Specifically, when the attacker injects an additional watermark with a different target label, it tends to induce a multi-backdoor conflict within the local model. This interference weakens the original backdoor and reduces the overall effectiveness of the attack. Overall, \texttt{Coward} remains resilient, as attackers struggle to maintain high ASR without compromising their own attack objectives.

\subsection{Why is Our Method Effective?}
\label{subsec:effectiveness}
To understand why our method works, we provide both overall and case-specific analyses on the CIFAR-10 dataset under a highly non-IID distribution with $\alpha = 0.3$.

\begin{figure}[t]
    \centering
    \includegraphics[width=0.42\linewidth]{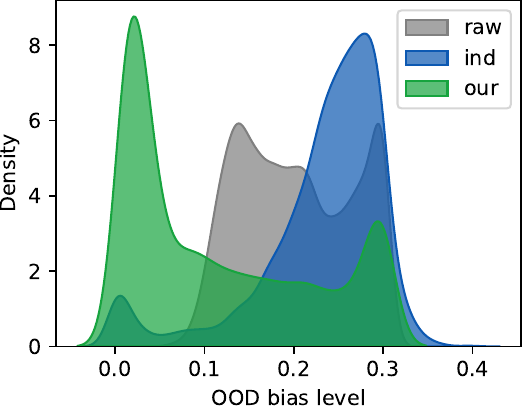}
    \includegraphics[width=0.44\linewidth]{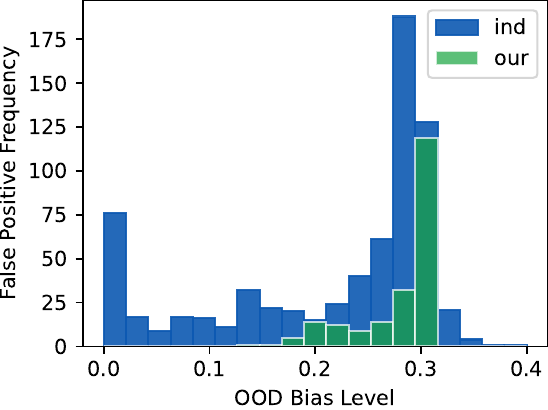}
    \vspace{-1.5em}
    \caption{\textbf{A Holistic Quantification of OOD Bias and Its Impact on FPR}. The left panel shows the bias distribution, while the right panel shows the relationship between bias severity and the number of misjudged benign clients. Our method significantly reduces the OOD bias level, leading to fewer misjudgments, whereas the Indicator increases the bias level and exhibits a high false positive frequency in biased regions.}
    \label{fig:overall_ood_bias_evidence}
\end{figure}

\vspace{0.3em}
\noindent \textbf{Overall OOD Bias Analysis}.
We evaluate overall bias under vanilla FL, BackdoorIndicator, and our method, and further examine its correlation with the detection false positive rate.
Specifically, the prediction bias is quantified by the standard deviation of class-wise prediction probabilities, since uniform predictions are ideal for OOD samples, and any bias toward a specific class results in non-uniformity, thereby increasing the variance.
By aggregating the bias across all participating clients over all training epochs, we obtain the overall bias level.
As the results shown in Figure~\ref{fig:overall_ood_bias_evidence} indicate, it is evident that: \textbf{(1)} \textit{Our method yields the lowest bias distribution}, validating the effectiveness of regulating the OOD mapping and highlighting the adverse effect of BackdoorIndicator's random mapping strategy. 
\textbf{(2)} \textit{Even at comparable bias levels, our method achieves a lower FPR}, demonstrating its robustness against OOD bias.

\begin{figure}[t]
    \centering
    \includegraphics[width=0.19\linewidth]{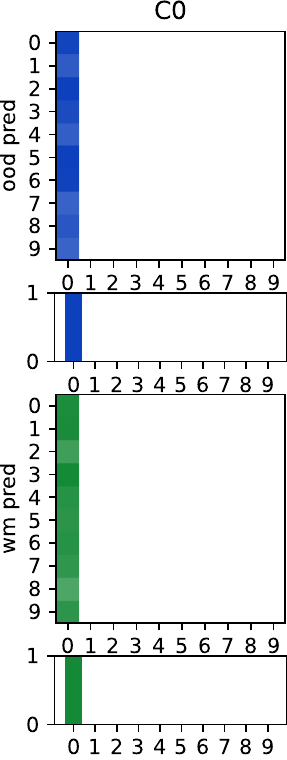}
    \includegraphics[width=0.19\linewidth]{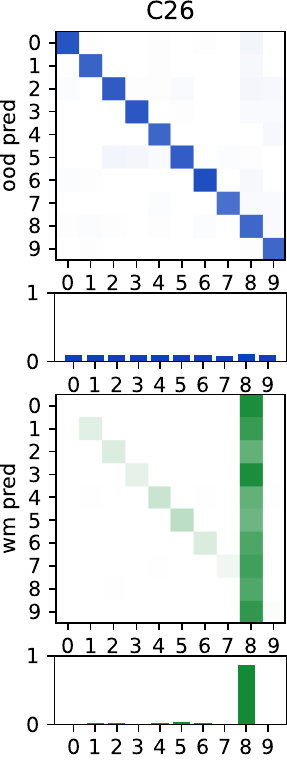}
    \includegraphics[width=0.19\linewidth]{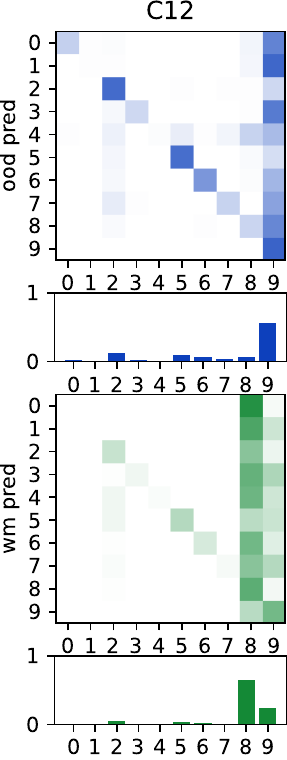}
    \includegraphics[width=0.19\linewidth]{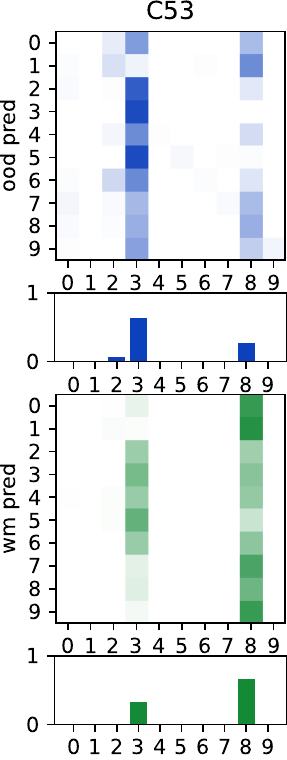}
    \includegraphics[width=0.19\linewidth]{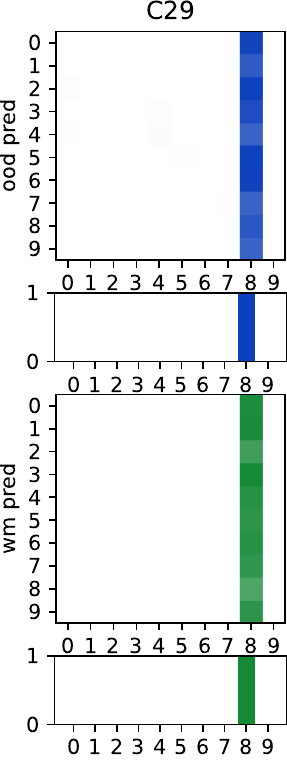}
    \vspace{-1.5em}
    \caption{\textbf{Case study of client detection with our method.} The first two columns illustrate ideal detection outcomes for malicious and benign clients, respectively. The remaining columns show successful identification of benign clients under different levels of OOD bias.}
    \label{fig:ood_robust_of_watermark}
\end{figure}

\vspace{0.3em}
\noindent \textbf{Case Study}.
To intuitively explain how our method mitigates OOD bias, we present a set of representative detection cases of our method in Figure~\ref{fig:ood_robust_of_watermark}, ranging from an ideal unbiased scenario to increasingly biased conditions.
Each case visualizes the prediction confusion matrices of both raw and watermarked OOD samples, revealing how OOD bias impact the two injected mappings.
Specifically, the top two rows show the prediction results of raw OOD samples, while the bottom rows show those of watermarked samples.
Each case includes a small subfigure illustrating the cumulative distribution across target classes.
To aid intuitive understanding, we highlight that the attack target is class 0, and the watermark target is class 8.
We elaborate on the case study by explaining both the ideal unbiased scenario and the biased scenario.
\textbf{\textit{(1) Ideal Cases:}} Case 1 illustrates a malicious client (C0), where both the OOD and watermark mappings collapse into the attack target class 0, indicating severe OOD mapping distortion and a strong collision effect caused by the backdoor.
This results in low watermark accuracy, enabling effective detection.
In contrast, Case 2 represents an ideal benign client unaffected by OOD bias, where both the OOD (diagonal) and watermark mappings are fully preserved.
Empirically, OOD bias seldom overrides backdoors but often impacts benign clients. 
\textbf{\textit{(2) Biased Cases:}} Case 3 presents a mild bias where the OOD mapping is largely retained but partially biased toward class 9, yet the watermark prediction remains dominated by class 8.
This highlights the strength of the watermark, which consistently guides triggered samples to the target class, even when raw samples are misclassified.
In Case 4, under a stronger bias where OOD predictions collapse into two dominant classes, the watermark still holds the highest confidence for class 8, indicating resilience even under severe distributional shifts.
Intuitively, after watermark planting, the model forms a direct association between the trigger pattern and the target label, making it less influenced by the background image and thus more robust to OOD bias.
Case 5 presents an extreme scenario where the OOD mapping collapses to a single class that coincidentally aligns with the watermark target.
\textit{These cases collectively indicate the robustness of our method against different degrees of OOD bias.}

\section{Conclusion}
\label{sec:conclusion}
In this paper, we revisited existing FL backdoor detection methods and identified two fundamental limitations: non-IID divergence and OOD prediction bias. To tackle these challenges, we designed a new method, \texttt{Coward}, which leverages a collision-based OOD watermark that the server injects and later inspects to expose malicious clients. Inspired by the \emph{multi-backdoor collision effects}, we carefully designed \texttt{Coward} to enable an inverted detection paradigm and to minimally interfere with benign pattern learning, thereby mitigating the severity of OOD prediction bias and its adverse effects on misjudgment. Extensive experiments verified the effectiveness of \texttt{Coward} and its resilience against potential adaptive attacks. 

\section*{Acknowledgement}
We sincerely thank Dr. Kangjie Chen and Prof. Dacheng Tao from Nanyang Technological University for their constructive suggestions on early drafts of this work.

\section*{Declaration of generative AI and AI-assisted technologies in the manuscript preparation process}
During the preparation of this work the authors used ChatGPT in order to improve language clarity and readability. After using this tool, the authors reviewed and edited the content as needed and take full responsibility for the content of the article.

\bibliographystyle{elsarticle-num}
\bibliography{main,FL}

\clearpage
\begingroup
\hypersetup{pageanchor=false}
\renewcommand{\theHfigure}{appendix.\arabic{figure}}
\renewcommand{\theHtable}{appendix.\arabic{table}}
\makeatletter
\@ifundefined{theHsubfigure}{}{\renewcommand{\theHsubfigure}{appendix.\arabic{figure}.\arabic{subfigure}}}
\makeatother
\begin{appendices}

\setcounter{page}{1}
\setcounter{table}{0}
\setcounter{figure}{0}

\begin{center}
\section*{\Large\bfseries Supplementary}  
\end{center}

\section{Collision Effects on OOD Watermark}
\label{appendix:collision_ood}
In Section~\ref{subsec:revisit_collision}, we introduced the multi-backdoor collision effect under a centralized setting where two backdoors share the same dataset, aiming to illustrate its core mechanism. To further demonstrate how this effect manifests in our OOD-based watermark under practical FL scenarios, we provide an extended analysis in this section. Specifically, we examine the collision effect between the attacker’s backdoor and our OOD watermark across progressively complex scenarios: starting from a centralized setup, moving to a static FL environment with fixed client participation, and finally to the dynamic FL scenario adopted in our main experiments. To align with the main experiments, we conduct all evaluations under the 0.9 non-i.i.d. setting on the CIFAR-10 dataset by default. As elaborated in the following sections, results across all three stages consistently reveal a clear collision effect, validating the capability of our OOD watermark to trigger such interactions and effectively distinguish malicious clients. 

\begin{figure}[!t]
\centering
\subfloat[\textit{w./} BN switch]{%
    \includegraphics[width=0.35\linewidth]{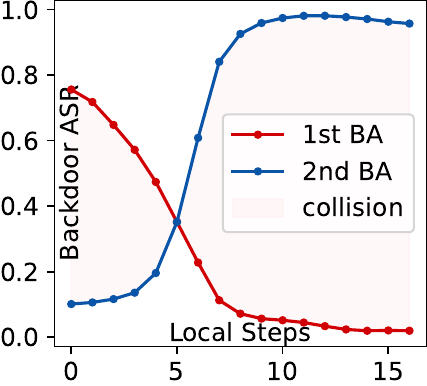}
    \includegraphics[width=0.35\linewidth]{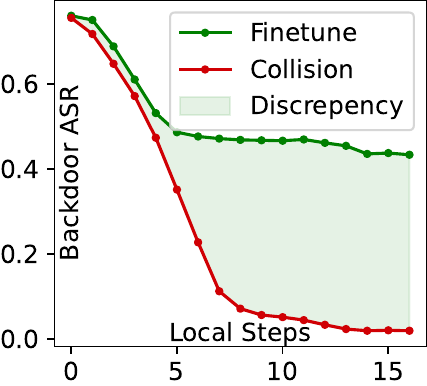}
    \label{fig:collision_ood_bn}
}
\hfil
\subfloat[\textit{w./o.} BN switch]{%
    \includegraphics[width=0.35\linewidth]{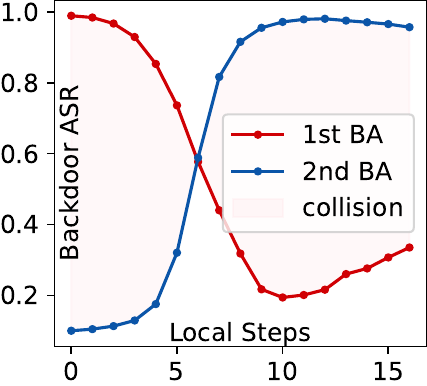}
    \includegraphics[width=0.35\linewidth]{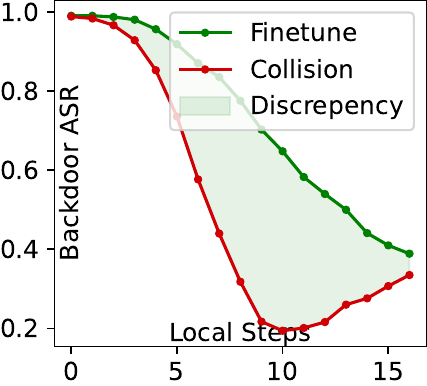}
    \label{fig:collision_ft_bn}
}

\caption{\textbf{OOD watermark collision under centralized scenario}. Our OOD watermark is planted as the second backdoor. The resulting collision effect is significant, regardless of whether the BN layer is switched. However, switching the BN layer creates a more pronounced performance discrepancy between benign fine-tuning and backdoor injection.}
\label{fig:collision_on_ood}
\end{figure}

\vspace{0.3em}
\noindent \textbf{Centralized Scenario}. We conduct a similar experiment as in Section~\ref{subsec:revisit_collision}, with the key difference that the second backdoor injection is replaced by our OOD watermark planting process. Specifically, we use a target label 8 watermark with MNIST as the planting set. As shown in Fig.~\ref{fig:collision_on_ood}, in each subfigure, the left side illustrates the collision effect, while the right side shows the discrepancy between benign (\ie, fine-tuning) and malicious (\ie, backdoor injection) operations. Specifically, in subfigure (a), we observe a clear collision effect on the OOD watermark, characterized by a clear drop in ASR and a notable distinction between benign and malicious clients. In subfigure (b), when the BN layer switch is disabled, the benign-malicious discrepancy becomes indistinguishable, indicating the necessity of BN isolation for effective detection.

\begin{figure}[t!]
  \centering
  \subfloat[\textit{w./o.} BN switch]{%
    \includegraphics[width=0.42\linewidth,
                     trim=3.4mm 4.2mm 3.7mm 0mm,clip]{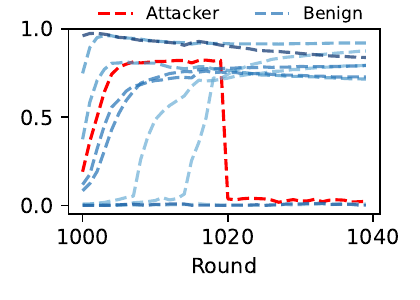}%
    \label{fig:no_bn}
  }
  \subfloat[\textit{w.} BN switch]{%
    \includegraphics[width=0.42\linewidth,
                     trim=3.4mm 4.2mm 3.7mm 0mm,clip]{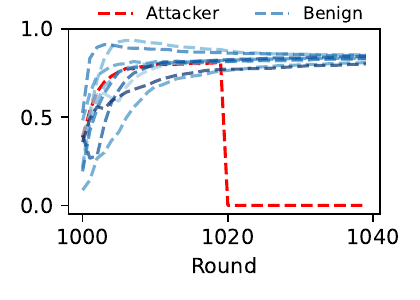}%
    \label{fig:with_bn}
  }
  \caption{\textbf{OOD watermark collision under static FL scenario}. Once the attacker (in \red{red}) begins injecting the backdoor at round 1021, its watermark accuracy rapidly drops to 0, indicating a strong collision effect and a clear distinction from benign clients. Moreover, the BN switch plays a critical role in accurately reflecting the benign clients' (in varying shades of \blue{blue}) ability to retain the watermark.}
  \label{fig:bn_effect}
\end{figure}

\vspace{0.3em}
\noindent \textbf{Static FL Scenario.}
We consider a slightly more complex scenario under a static FL setting, involving 10 clients with full participation in each communication round. Among them, only one client is malicious. This simplified full-participation setup facilitates the observation of each client's temporal behavior and helps illustrate the collision effect in a federated learning context. In each global round, the server injects an OOD watermark, while all benign clients perform standard local training. To clearly observe the collision effect, the attacker is configured to behave benignly for the first 20 rounds and initiates a backdoor attack targeting label 0 from round 21 onward. After each local update, we inspect the watermark accuracy of all 10 participating clients. As shown in Figure~\ref{fig:bn_effect}(b), once the malicious client begins the attack, its watermark accuracy drops sharply to nearly zero, indicating a strong collision effect. In contrast, the remaining nine benign clients consistently maintain high watermark accuracy. This sharp divergence highlights the benign-malicious discrepancy induced by the collision effect, which is critical for enabling effective backdoor detection. Additionally, in subfigure (a), we observe that benign clients struggle to preserve the watermark without BN layer isolation, underscoring the importance of BN switching in reducing FPR.

\begin{figure*}[t]
\vspace{-1em}
    \centering
    \includegraphics[width=0.98\linewidth]{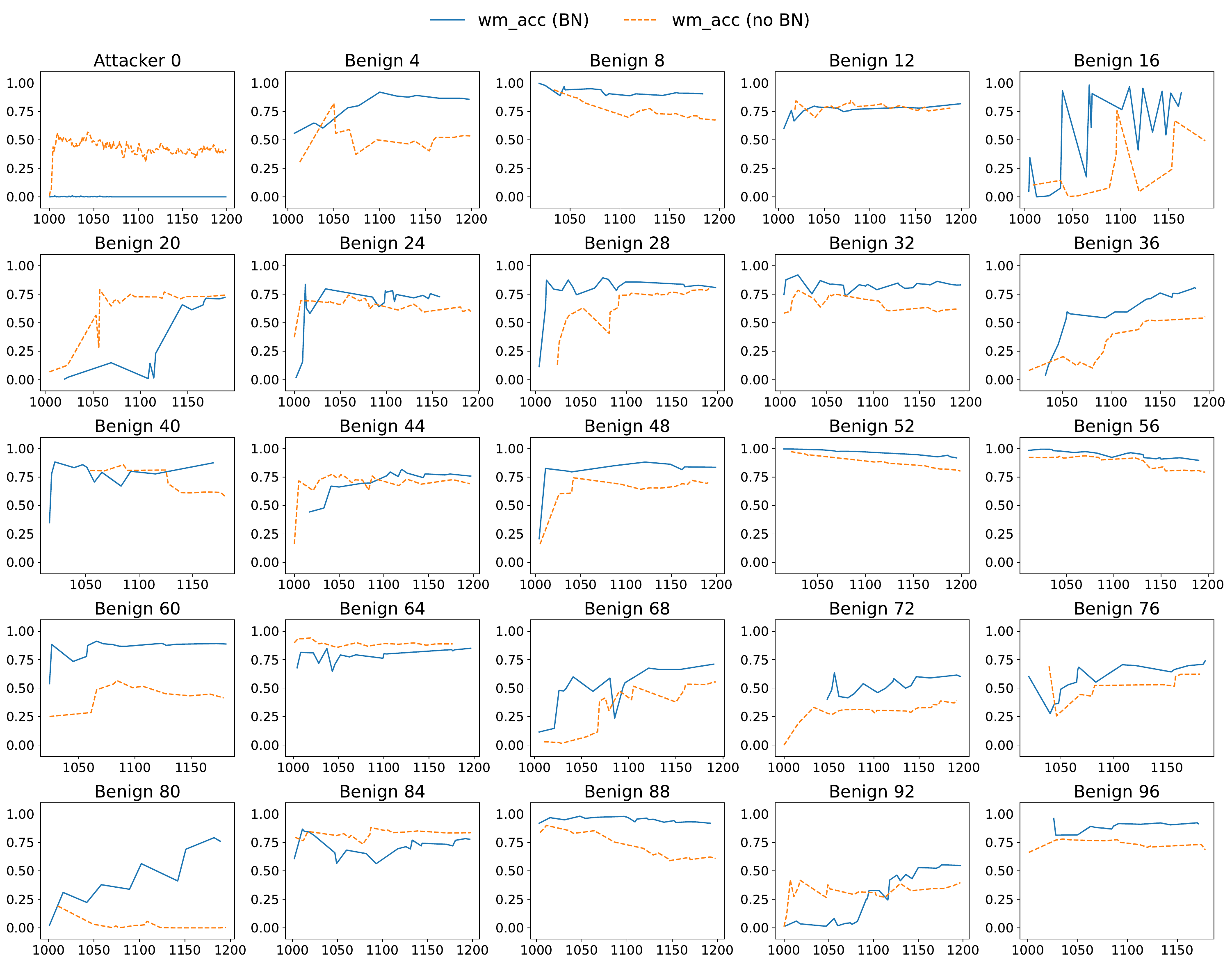}
    \caption{\textbf{OOD watermark collision under dynamic FL scenario.} The collision effect remains highly effective in distinguishing malicious behavior under dynamic federated participation. The attacker exhibits a strong collision effect, while benign clients show diverse but generally higher levels of watermark retention. 
    }
    \label{fig:wm_all_client}
    \vspace{-0.3cm}
\end{figure*}

\vspace{0.3em}
\noindent \textbf{Dynamic FL Scenario.}
We finally present the collision effect under the dynamic FL setting used in our main experiments, where 10 out of 100 clients are randomly selected to participate in each global round. This dynamic participation leads to divergent and inconsistent watermark retention behaviors among benign clients, making direct comparisons between them infeasible. Therefore, we focus on observing relative trends across rounds. Specifically, we plot the temporal variation of watermark accuracy for all clients based on their actual participation order. To control page length, we select clients at intervals of 4 (\ie, client IDs with a gap of 4). As shown in Figure~\ref{fig:wm_all_client}, we observe the following: \textbf{(1)} Malicious client 0 exhibits a pronounced collision effect, with watermark accuracy dropping to nearly zero. \textbf{(2)} Benign clients maintain relatively high watermark accuracy (mostly above 50\%), though the magnitude and variation patterns differ across clients due to heterogeneous participation and client distribution. Overall, the results reveal a clear separation between benign and malicious clients, enabling robust and tolerant threshold selection. Despite being influenced by dynamic participation and non-i.i.d. data, the collision effect remains effective in detection.

\begin{figure*}[t]
\vspace{-1em}
  \centering
\subfloat[attacker triggers]{%
\includegraphics[width=0.15\linewidth]{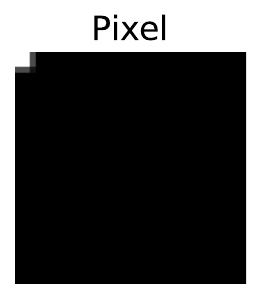}
\includegraphics[width=0.15\linewidth]{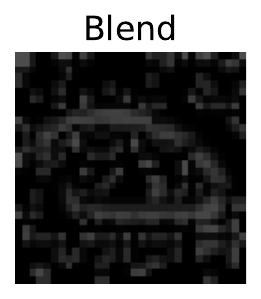}
\includegraphics[width=0.15\linewidth]{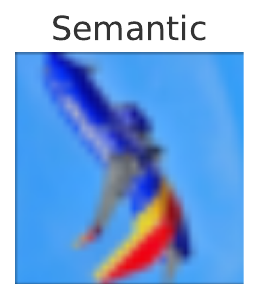}
}

\subfloat[watermark triggers]{%
\includegraphics[width=0.15\linewidth]{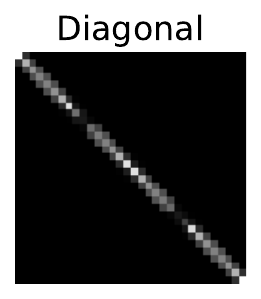}
\includegraphics[width=0.15\linewidth]{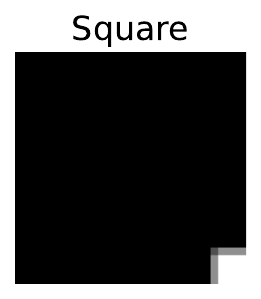}
\includegraphics[width=0.15\linewidth]{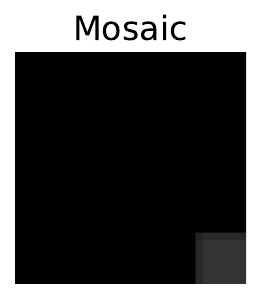}  
\includegraphics[width=0.15\linewidth]{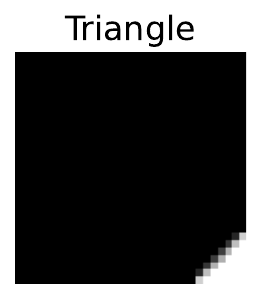}
\includegraphics[width=0.15\linewidth]{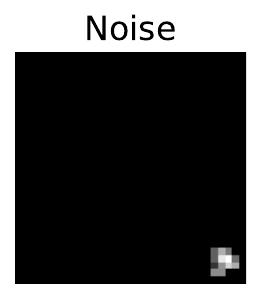}
\includegraphics[width=0.15\linewidth]{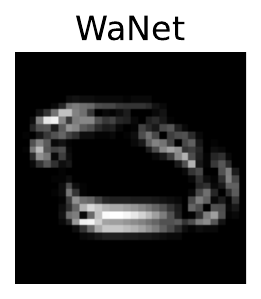}
}

\caption{Visualization of trigger configurations.}
\label{fig:trigger_pairs}
\end{figure*}

\begin{table}[t]
  \setlength{\abovecaptionskip}{0pt}
  \centering
  \caption{Performance of \texttt{Coward} with different OOD mappings on the CIFAR-10 dataset.}
  \vspace{0.1cm}
  \setlength{\tabcolsep}{1pt}
  \renewcommand{\arraystretch}{0.8}
  \begin{adjustbox}{max width=0.9\linewidth}
    \begin{tabular}{c|cccc}
      \toprule
      \text{OOD Mapping} & \text{TPR}$\uparrow$ & \text{FPR}$\downarrow$ & \text{ASR}$\downarrow$ & \text{BA}$\uparrow$ \\
      \midrule
      Default & 100.0 &  2.4 &  9.8 & 91.3 \\
      Shift+1 & 99.5 &  3.0 &  9.8 & 91.5 \\
      Shift+3 & 99.5 & 2.9 & 9.8 & 91.5 \\
      Shift+5 & 100.0 & 2.6 &  9.7 & 91.3 \\
      \bottomrule
    \end{tabular}
  \end{adjustbox}
  \label{tab:mapping}
\end{table}

\begin{table}[t]
  \setlength{\abovecaptionskip}{0pt}
  \centering
  \caption{Performance of \texttt{Coward} for all watermark labels.}
  \vspace{0.1cm}
  \setlength{\tabcolsep}{1pt}
  \renewcommand{\arraystretch}{0.8}
  \begin{adjustbox}{max width=0.6\linewidth}
    \begin{tabular}{c|cccc}
      \toprule
      \text{Label ($y_m$)} & \text{TPR}$\uparrow$ & \text{FPR}$\downarrow$ & \text{ASR}$\downarrow$ & \text{BA}$\uparrow$ \\
      \midrule
      1 & 100.0 &  9.2 &  9.9 & 91.4 \\
      2 & 100.0 &  6.5 &  9.8 & 91.5 \\
      3 & 100.0 & 10.4 & 10.0 & 91.5 \\
      4 & 100.0 &  5.1 &  9.8 & 91.5 \\
      5 & 100.0 & 10.1 &  9.9 & 91.2 \\
      6 & 100.0 &  5.6 &  9.9 & 91.5 \\
      7 & 100.0 & 13.6 &  9.7 & 91.5 \\
      8 &  98.5 &  4.0 & 10.2 & 91.6 \\
      9 & 100.0 & 17.8 &  9.7 & 91.4 \\
      \bottomrule
    \end{tabular}
  \end{adjustbox}
  \label{tab:wm_label}
\end{table}

\section{Discussion on Watermark Configuration}
\label{appendix:wm_cfg}
Our detection paradigm is flexible with respect to specific watermark configurations, allowing practical deployment scenarios to select or design their own. To concretely demonstrate this flexibility, we provide empirical evidence by evaluating constituent components of our OOD watermark: the OOD base mapping, the watermark trigger, and the watermark target. To align with the main experiments, we conduct all evaluations under the 0.9 non-i.i.d. setting on the CIFAR-10 dataset by default.

\vspace{0.3em}
\noindent\textbf{OOD Base Mapping.} The key role of the OOD base mapping is to provide a fixed foundation for building the watermark mapping. As long as the mapping remains class-wise and fixed, the specific choice of mapping is not critical. While we adopt a diagonal mapping in the main experiments for simplicity, we demonstrate here that our approach is compatible with other mapping configurations, highlighting its flexibility. Specifically, we replace the default diagonal mapping (\ie, \(i \rightarrow i\)) with three alternative 1-to-1 shift-based mappings: \(i \rightarrow (i{+}3) \bmod 10\), \(i \rightarrow (i{+}5) \bmod 10\), and \(i \rightarrow (i{+}7) \bmod 10\). As shown in Table~\ref{tab:mapping}, all variants achieve consistently strong detection performance, indicating that \texttt{Coward} does not rely on a specific OOD mapping rule. These results confirm that our method is robust to the choice of OOD assignment and can be seamlessly adapted to different mapping schemes.

\vspace{0.3em}
\noindent\textbf{Trigger Type.}
Previously, we have already presented quantitative results demonstrating the general effectiveness of the collision effect and the corresponding detection mechanism across various trigger pairs in Table~\ref{tab:triggerpair}. To provide a more intuitive understanding, we further visualize the appearance of the triggers. As illustrated in Figure~\ref{fig:trigger_pairs}, our server-malicious trigger pairs include both semantically similar combinations (\eg, a server-side square pattern and a client-side pixel-style trigger) and semantically dissimilar ones (\eg, a server-side WaNet trigger and a client-side semantic trigger). Notably, both types consistently yield strong detection performance. These results indicate that the effectiveness of our approach is largely independent of the visual semantics of the triggers, offering defenders a broad design space to customize trigger appearances according to their specific needs.

\vspace{0.3em}
\noindent\textbf{Watermark Label.}
In the main experiments, we fixed the watermark target label to 8 for simplicity. However, we emphasize that the effectiveness of \texttt{Coward} is generally applicable to any target label. To validate this flexibility, we present results in Table~\ref{tab:wm_label}, where the attack target is fixed to label 0 and the watermark label is varied across all possible values. The consistently strong detection performance across all watermark labels confirms the adaptability of our approach in selecting the watermark target.

\begin{table*}[t]
\setlength{\abovecaptionskip}{0pt}
\centering
\caption{Ablation study of \texttt{Coward} on watermark injection strength.}
\label{tab:wm_strength}
\vspace{0.2cm}

\begin{minipage}{0.45\textwidth}
\centering
\setlength{\tabcolsep}{1pt}
\renewcommand{\arraystretch}{0.8}
\begin{adjustbox}{max width=\linewidth}
\begin{tabular}{c|cccc}
  \toprule
  \text{Size} & \text{TPR}$\uparrow$ & \text{FPR}$\downarrow$ & \text{ASR}$\downarrow$ & \text{BA}$\uparrow$ \\
  \midrule
  200 & 100.0& 23.7 & 10.0  & 91.5   \\
  500 & 98.0 & 2.2 & 10.1 & 91.6 \\
  1000 & 99.0 & 4.2 & 10.1 & 91.4 \\
  1500 & 100.0 & 3.1 & 10.1 & 91.1 \\
  2000 & 100.0 & 3.3 & 10.1 & 91.4 \\
  \bottomrule
\end{tabular}
\end{adjustbox}
\vspace{0.1cm}

{\small\textbf{(a)} Varying size of planting set\par}

\end{minipage}
\hfill
\begin{minipage}{0.48\textwidth}
\centering
\setlength{\tabcolsep}{1pt}
\renewcommand{\arraystretch}{0.8}
\begin{adjustbox}{max width=\linewidth}
\begin{tabular}{c|cccc}
  \toprule
  \text{wm\_lr} & \text{TPR}$\uparrow$ & \text{FPR}$\downarrow$ & \text{ASR}$\downarrow$ & \text{BA}$\uparrow$ \\
  \midrule
  0.0001 & 100.0 & 7.1 & 10.1 & 91.5 \\
  0.0005 & 99.0 & 5.6 & 10.0 & 91.6 \\
  0.001 & 100.0 & 3.3 & 9.8 & 91.4 \\
  0.01 & 98.0 & 1.8 & 9.9 & 91.5 \\
  \bottomrule
\end{tabular}
\end{adjustbox}

\vspace{0.05cm}

{\small\textbf{(b)} Varying watermark learning rate\par}
\end{minipage}

\vspace{0.25cm}

\begin{minipage}{0.48\textwidth}
\centering
\setlength{\tabcolsep}{1pt}
\renewcommand{\arraystretch}{0.8}
\begin{adjustbox}{max width=\linewidth}
\begin{tabular}{c|cccc}
  \toprule
  \text{wm\_rds} & \text{TPR}$\uparrow$ & \text{FPR}$\downarrow$ & \text{ASR}$\downarrow$ & \text{BA}$\uparrow$ \\
  \midrule
  1 & 100.0 & 11.6 & 10.2 & 91.6 \\
  2 & 100.0 & 6.4 & 9.9 & 91.3 \\
  5 & 100.0 & 3.3 & 9.8 & 91.4 \\
  10 & 100.0 & 5.7 & 10.0 & 91.4 \\
  20 & 100.0 & 4.2 & 9.9 & 91.2 \\
  \bottomrule
\end{tabular}
\end{adjustbox}

\vspace{0.2cm}

{\small\textbf{(c)} Varying injection rounds\par}
\end{minipage}

\vspace{-0.4cm}
\end{table*}

\begin{table}[t]
\centering
\caption{Performance of \texttt{Coward} under different levels of global model regularization.}\vspace{0.2cm}

\setlength{\tabcolsep}{1pt}
\renewcommand{\arraystretch}{0.8}
\begin{tabular}{c|cccc}
  \toprule
  \textbf{$\lambda$} & \text{TPR}$\uparrow$ & \text{FPR}$\downarrow$ & \text{ASR}$\downarrow$ & \text{BA}$\uparrow$ \\
  \midrule
    0 & 93.5 & 6.5 & 10.3 & 90.4 \\
    0.1 & 96.0 & 6.8 & 10.2 & 90.1 \\
    0.3 & 99.0 & 10.9 & 10.3 & 90.4 \\
    0.5 & 97.0 & 9.6 & 10.2 & 90.4 \\
  \bottomrule
\end{tabular}

\label{tab:gm_reg}
\vspace{-0.4cm}
\end{table}

\section{Ablation Study on Watermark Injection}
\label{appendix:wm_inj}

Beyond the design choices of the watermark itself, an equally important yet distinct problem is how to effectively implant the watermark. In this section, we conduct ablation studies on all factors that influence watermark injection. To align with the main experiments, we conduct all evaluations under the 0.9 non-IID setting on the CIFAR-10 dataset by default.

\vspace{0.3em}
\noindent \textbf{Impact of Global Model Regularization.}
Table~\ref{tab:gm_reg} reports the effect of varying levels of global model regularization. We observe a positive impact on TPR, where applying regularization consistently yields higher TPRs compared to the no-regularization baseline. Increasing the regularization strength $\lambda$ leads to slight improvements in TPR, with the best performance observed at $\lambda = 0.3$. These results suggest that a moderate level of regularization is beneficial for achieving optimal detection performance. 

\vspace{0.3em}    
\noindent \textbf{Learning Efficiency of Watermark Injection}. 
We conduct ablation studies on three key hyperparameters that control the strength of watermark injection: the number of OOD watermark samples, the watermark learning rate, and the number of injection iterations per global round. We vary one factor at a time while keeping the others fixed. By default, the watermark learning rate is set to 0.001, the number of watermark injection rounds is set to 5, and the watermark dataset size is set to 1000. As shown in Table~\ref{tab:wm_strength},
\texttt{Coward} consistently achieves strong performance, with TPR exceeding 98\% and FPR remaining below 10\% in most cases. Even under extremely constrained injection settings (\eg, a watermark dataset size of 200, a learning rate of 0.0001, or a single injection round), the method still maintains competitive results, achieving 100\% TPR with FPR below 25\%. Overall, effective watermark injection requires only a low level of effort. Most parameter variations lead to only marginal performance differences, and any configuration beyond the most extreme cases yields satisfactory performance. These results indicate that our method is robust across a wide range of watermark injection strengths and introduces negligible computational overhead in federated learning. \textit{Notably, effective defense can be achieved with as few as 1 round of watermark injection}. 

\begin{table}[t]
\caption{Attack success rate and benign accuracy against stealthy single-client backdoor attacks with advanced training strategies. Result highlighting follows the same rules as Table~\ref{tab:noniid}.}
\label{tab:mainexp_asr_ba}
\centering
\scriptsize
\setlength{\tabcolsep}{1pt}
\renewcommand{\arraystretch}{1.1}
\begin{adjustbox}{width=\textwidth}
\begin{tabular}{cc|*{14}{c}}
\toprule
\multirow{2}{*}{\text{Dataset}} & \multirow{2}{*}{\text{Attack}} & \multicolumn{2}{c}{\text{MultiKrum}} & \multicolumn{2}{c}{\text{FoolsGold}} & \multicolumn{2}{c}{\text{Rflbat}} & \multicolumn{2}{c}{\text{DeepSight}} & \multicolumn{2}{c}{\text{Flame}} & \multicolumn{2}{c}{\text{Indicator}} & \multicolumn{2}{c}{\textbf{\textbf{Ours}}} \\
\cmidrule(lr){3-4}\cmidrule(lr){5-6}\cmidrule(lr){7-8}\cmidrule(lr){9-10}\cmidrule(lr){11-12}\cmidrule(lr){13-14}\cmidrule(lr){15-16}
 &  & ASR$\downarrow$ & BA$\uparrow$ & ASR$\downarrow$ & BA$\uparrow$ & ASR$\downarrow$ & BA$\uparrow$ & ASR$\downarrow$ & BA$\uparrow$ & ASR$\downarrow$ & BA$\uparrow$ & ASR$\downarrow$ & BA$\uparrow$ & ASR$\downarrow$ & BA$\uparrow$ \\
\midrule
\multirow{4}{*}{\text{CIFAR10}} & \text{Vanilla} & \textcolor{red}{97.0} & 91.7 & \red{92.8} & \textbf{91.9} & \red{96.6} & \textbf{91.9} & 30.2 & 91.8 & \textcolor{red}{96.3} & 91.7 & 4.9 & \textbf{91.9} & \textbf{4.8} & 91.5 \\
 & \text{PGD} & \textcolor{red}{97.3} & 91.7 & \textcolor{red}{92.7} & \textbf{91.9} & \textcolor{red}{96.4} & 91.8 & 32.3 & \textbf{91.9} & \textcolor{red}{97.8} & 91.7 & \textbf{5.5} & 91.6 & 5.9 & 91.4 \\
 & \text{NTX} & \textcolor{red}{96.4} & 91.7 & \textcolor{red}{96.0} & 91.8 & \textcolor{red}{95.5} & 91.8 & 41.1 & \textbf{91.9} & \textcolor{red}{95.5} & 91.4 & \textcolor{red}{79.1} & 91.4 & \textbf{4.1} & 91.6 \\
 & \text{CML} & \textcolor{red}{94.1} & \textbf{91.8} & \textcolor{red}{88.4} & 91.7 & \textcolor{red}{93.4} & \textbf{91.8} & \textcolor{red}{62.4} & 91.3 & \textcolor{red}{94.9} & 91.6 & \textbf{8.4} & 91.3 & 26.1 & 91.4 \\
\midrule
\multirow{4}{*}{\text{CIFAR100}} & \text{Vanilla} & \textcolor{red}{92.9} & 69.5 & \textcolor{red}{88.9} & \textbf{69.6} & \textcolor{red}{92.8} & \textbf{69.6} & \textcolor{red}{53.1} & 69.4 & \textcolor{red}{93.0} & 69.4 & 9.3 & 68.9 & \textbf{0.8} & 68.8 \\
 & \text{PGD} & \textcolor{red}{92.9} & 69.6 & \textcolor{red}{92.0} & \textbf{69.7} & \textcolor{red}{92.7} & \textbf{69.7} & 49.3 & 69.6 & \textcolor{red}{93.4} & 69.4 & 16.1 & 68.9 & \textbf{0.8} & 69.0 \\
 & \text{NTX} & \textcolor{red}{91.4} & 69.4 & \textcolor{red}{91.7} & 69.4 & \textcolor{red}{92.7} & \textbf{69.7} & 47.4 & 69.6 & \textcolor{red}{92.9} & 69.3 & 2.3 & 69.2 & \textbf{0.8} & 69.0 \\
 & \text{CML} & \red{85.6} & 69.3 & \red{90.7} & 69.5 & \red{86.3} & \textbf{69.6} & \red{60.7} & 69.4 & \red{87.1} & 69.4 & 16.0 & 66.7 & \textbf{0.9} & 69.1 \\
\midrule
\multirow{4}{*}{\text{EMNIST}} & \text{Vanilla} & \red{100.0} & \textbf{99.7} & \red{100.0} & \textbf{99.7} & \red{100.0} & \textbf{99.7} & \red{100.0} & \textbf{99.7} & \red{100.0} & \textbf{99.7} & {21.4} & 99.7 & \textbf{0.0} & \textbf{99.7} \\
 & \text{PGD} & \red{99.9} & \textbf{99.7} & \red{100.0} & \textbf{99.7} & \red{100.0} & \textbf{99.7} & \red{99.8} & \textbf{99.7} & 33.0 & \textbf{99.7} & 25.3 & 99.6 & \textbf{0.0} & \textbf{99.7} \\
 & \text{NTX} & \red{99.9} & \textbf{99.7} & \red{100.0} & \textbf{99.7} & \red{100.0} & \textbf{99.7} & \red{100.0} & \textbf{99.7} & \red{96.8} & \textbf{99.7} & {16.0} & 98.9 & \textbf{0.1} & \textbf{99.7} \\
 & \text{CML} & \red{100.0} & \textbf{99.7} & \red{100.0} & \textbf{99.7} & \red{100.0} & 99.6 & \red{100.0} & \textbf{99.7} & 29.3 & \textbf{99.7} & {39.2} & 99.6 & \textbf{15.4} & \textbf{99.7} \\
\bottomrule
\end{tabular}
\end{adjustbox}
\end{table}

\begin{table}[t]
\caption{Attack success rate and benign accuracy under multi-client backdoor attacks on the CIFAR-10 dataset. Result highlighting follows the same rules as Table~\ref{tab:noniid}.}
\label{tab:multi_asr_ba}
\centering
\scriptsize
\setlength{\tabcolsep}{1pt}
\begin{adjustbox}{width=\textwidth}

\begin{tabular}{cc|*{14}{c}}
\toprule
\multirow{2}{*}{\text{Strategy}} & \multirow{2}{*}{\text{Atk\%}} & \multicolumn{2}{c}{\text{MultiKrum}} & \multicolumn{2}{c}{\text{FoolsGold}} & \multicolumn{2}{c}{\text{Rflbat}} & \multicolumn{2}{c}{\text{DeepSight}} & \multicolumn{2}{c}{\text{Flame}} & \multicolumn{2}{c}{\text{Indicator}} & \multicolumn{2}{c}{\textbf{\textbf{Coward}}} \\
\cmidrule(lr){3-4}\cmidrule(lr){5-6}\cmidrule(lr){7-8}\cmidrule(lr){9-10}\cmidrule(lr){11-12}\cmidrule(lr){13-14}\cmidrule(lr){15-16}
 &  & ASR$\downarrow$ & BA$\uparrow$ & ASR$\downarrow$ & BA$\uparrow$ & ASR$\downarrow$ & BA$\uparrow$ & ASR$\downarrow$ & BA$\uparrow$ & ASR$\downarrow$ & BA$\uparrow$ & ASR$\downarrow$ & BA$\uparrow$ & ASR$\downarrow$ & BA$\uparrow$ \\
\midrule
\multirow{3}{*}{\text{Uniform}} & \text{30\%} & 10.0 & 91.6 & \textbf{9.7} & \textbf{91.8} & \textcolor{red}{88.7} & 91.6 & \textcolor{red}{97.8} & 91.6 & \textbf{9.7} & \textbf{91.9} & 11.1 & 91.5 & 9.9 & 91.5 \\
 & \text{50\%} & \red{98.4} & 90.8 & 10.1 & \textbf{91.8} & \red{91.9} & 91.4 & \red{98.2} & 91.4 & \red{99.2} & 91.4 & 12.2 & 91.2 & \textbf{9.9} & 91.2 \\
 & \text{70\%} & \red{98.2} & 90.8 & \textbf{9.7} & \textbf{91.8} & \red{93.5} & 90.0 & \red{98.8} & 91.5 & \red{99.0} & 91.5 & 40.5 & 88.8 & 10.5 & 89.6 \\
\midrule
\multirow{3}{*}{\text{DBA}} & \text{30\%} & 9.8 & 91.6 & \textbf{9.5} & \textbf{91.9} & \textcolor{red}{81.9} & 91.5 & \textcolor{red}{90.6} & 91.5 & \textbf{9.5} & \textbf{91.9} & 10.1 & 91.8 & 9.7 & 91.3 \\
 & \text{50\%} & \red{96.7} & 91.1 & \textbf{9.9} & 91.9 & \red{90.7} & 91.3 & \red{94.7} & 91.6 & \red{81.9} & 91.7 & 10.6 & 91.5 & 10.0 & 91.2 \\
 & \text{70\%} & \red{97.7} & 90.1 & \textbf{9.6} & \textbf{91.8} & \textcolor{red}{95.3} & 88.8 & \textcolor{red}{96.0} & 90.9 & \textcolor{red}{96.5} & 91.3 & 11.9 & 89.4 & 10.3 & 89.6 \\
\midrule
\multirow{3}{*}{\text{NBA}} & \text{30\%} & \textbf{10.1} & 91.7 & \textcolor{red}{88.6} & 91.7 & \textcolor{red}{65.7} & 91.7 & \textcolor{red}{90.1} & 91.8 & \textcolor{red}{86.5} & \textbf{92.0} & 11.0 & 91.2 & \textbf{10.1} & 91.5 \\
 & \text{50\%} & \textcolor{red}{91.5} & 91.6 & \textcolor{red}{93.4} & 91.6 & \textcolor{red}{85.2} & 91.4 & \textcolor{red}{93.9} & \textbf{91.7} & \textcolor{red}{94.8} & 91.1 & 32.6 & 91.5 & \textbf{10.1} & 91.5 \\
 & \text{70\%} & \textcolor{red}{96.5} & 90.4 & \textcolor{red}{96.1} & 90.8 & \textcolor{red}{92.1} & 89.4 & \textcolor{red}{97.1} & 90.6 & \textcolor{red}{96.4} & 90.6 & \textcolor{red}{66.7} & 90.8 & \textbf{24.7} & \textbf{91.6} \\
\bottomrule
\end{tabular}
\end{adjustbox}
\end{table}

\section{Additional ASR and BA Results}
\label{appendix:asr_ba}
We provide the attack success rate (ASR) and benign accuracy (BA) results for the advanced single-attacker and multi-attacker experiments, respectively.

\section{Potential Limitations and Future Directions}
While our method provides a simple and effective early solution for proactive defense, several open directions remain worth further exploration and refinement in future work on this topic:

\begin{itemize}[leftmargin=*]
    \item \textbf{Fine-grained OOD Watermark Design}. Our current OOD watermark adopts a simple yet effective configuration applied uniformly across the entire planting set, using widely adopted trigger patterns. Exploring more complex and fine-grained designs that potentially leverage the diversity of OOD samples (\eg, sample-specific or class-conditional watermark mappings) would be a valuable future direction.
    
    \item \textbf{Dynamic Watermarking Strategy}. In the current version, a fixed watermark configuration is used throughout training, enabling effective injection and maintaining a consistent collision signal. Nevertheless, our framework is also flexible enough to support dynamic watermark configurations (\ie, across training rounds or even client-specific setups). This adaptability, when paired with carefully designed injection methods, would open up a new direction for enhancing the robustness and flexibility of the defense strategy.
    
    \item \textbf{Multi-Metric and Multi-Mechanism Detection}. While our method has already achieved effective detection using a single metric (\ie, watermark accuracy) with clear and consistent signals, the broader flexibility of the proposed framework allows for further expansion. Specifically, it can support joint observation of client responses to multiple operations, opening up the possibility of incorporating diverse detection mechanisms. For example, one may simultaneously leverage the maintenance and collision effects by separately monitoring OOD mappings and watermark mappings. This enriched perspective could provide complementary signals and represents a promising future direction.
\end{itemize}

\end{appendices}

\clearpage
\endgroup

\end{document}